\documentclass[aps,pre,reprint,groupedaddress]{revtex4-2}

\usepackage{amsmath}
\usepackage{amssymb}
\usepackage{graphicx}
\usepackage{xcolor}
\usepackage[normalem]{ulem}

\begin{document}


\title{The Dynamics of Fluctuating Thin Sheets Under Random Forcing}


\author{Chanania Steinbock}
\author{Eytan Katzav}
\affiliation{Racah Institute of Physics, The Hebrew University, Jerusalem 9190401, Israel}


\date{\today}

\begin{abstract}
We study the dynamic structure factor of fluctuating elastic thin sheets subject to conservative (athermal) random forcing. In Steinbock, Katzav \& Boudaoud, \textit{Phys. Rev. Research} \textbf{4}, 033096 (2022), the static structure factor of such a sheet was studied. In this paper, we recap the model developed there and investigate its dynamic properties. Using the self-consistent expansion (SCE), the time dependent two-point function of the height profile is determined and found to decay exponentially in time. Despite strong nonlinear coupling, the decay rate of the dynamic structure factor is found to coincide with the effective coupling constant for the static properties which suggests that the model under investigation exhibits certain quasi-linear behaviour. Confirmation of these results by numerical simulations is also presented.
\end{abstract}


\maketitle

\section{Introduction}
\label{sec:intro}

Thin sheets and surfaces are ubiquitous in everyday life yet the theory of their physical properties remains incomplete. For instance, despite the fact that crumpled paper balls take but a moment to make, the response of a thin sheet to random forcing remains poorly understood. Since randomly driven thin surfaces are relevant to a wide diversity of fields, ranging from the physics of crumpled paper to the properties of graphene to the behaviour of cell membranes, a theory of randomly driven surfaces derived from first principles would have far reaching consequences.

Loosely speaking, we can distinguish between two types of random forcing, completely uncorrelated white noise typical of thermal fluctuations and deliberate correlated noise such as the type of forcing applied when crumpling a sheet of paper. Here, we will focus on the latter kind of noise. Perhaps the easiest way to probe the structure of an athermally fluctuating sheet is to study the properties of the resultant crumpled sheet and indeed this is an active field of research in its own right, both experimentally \cite{PlouraboueRoux1996, Matan2002, BlairKudrolli2005, Balankin2006, Andresen2007, Balankin2008, Deboeuf2013, Balankin2013, Lahini2017, Gottesman2018, Shohat2022} and through mathematical or numerical modeling \cite{PlouraboueRoux1996, Vliegenthart2006, Sultan2006, Andrejevic2021}. The development of the theory of singular structures supported by thin sheets such as d-cones and ridges \cite{Lobkovsky1995, Lobkovsky1997, BenAmar1997, Chaieb1997, Cerda1999, Mora2002, Liang2005} has gone some way to bridging research into crumpled sheets with that of fluctuating sheets however its applicability has been limited by the impracticality of characterising sheets with more than a handful of ridges. Additionally, the structure of crumpled sheets can at most inform us of the static unvarying properties of fluctuating systems. To obtain insight into the complete dynamic structure of a fluctuating thin sheet, we must tackle such a system directly.

Previous research into the time-dependent dynamic structure of fluctuating surfaces is limited but has been studied in the context of tethered surfaces \cite{Kantor1986, Kantor1987} and polymerised membranes \cite{Frey1991, Niel1989}, though this research focused exclusively on thermal driving by white noise. In the context of tethered surfaces \cite{Kantor1986, Kantor1987}, the dynamics of phantom and self-avoiding flexible sheets was studied though at the cost of neglecting the elastic properties of real sheets. In contrast, \cite{Frey1991} focuses on the dynamic character of elastic polymerised membranes coupled to a random perturbing flowing fluid. Finally, \cite{Niel1989} uses a super-symmetric $\varepsilon$-expansion of a $D=4-\varepsilon$ dimensional membrane to obtain the dynamic exponent of an elastic thermally fluctuating polymerised membrane.

Recently, we showed how the static properties of a fluctuating sheet can be derived directly by applying techniques from out-of-equilibrium statistical mechanics to the physics of elastic systems \cite{Steinbock2022}. In particular, we developed a dynamic variant of the F\"oppl-von K\'arm\'an equations which describes the deformations of thin sheets and used this to obtain the static structure factor of a fluctuating thin sheet driven by athermal noise. In the current paper, we extend this approach to derive the time-dependent structure factor of the fluctuating sheet. This dynamic structure should be of direct relevance in understanding many features of the sheet, including its acoustic emissions \cite{Kramer1996,Houle1996,Sethna2001, Mendes2010}, optical signature \cite{Rad2019} and dissipative character \cite{Mehreganian2019, Mehreganian2021}. Further applications to diverse fields \cite{NelsonBook2004} such as the biophysics of cell membranes \cite{Frey1991, Liang2016}, the properties and stability of fluctuating graphene sheets \cite{Meyer2007, Meyer2007a, Fasolino2007, Thompson2009, Deng2016, Ahmadpoor2017} and wave turbulence \cite{Hassaini2019} can also be envisioned.

The paper is organised as follows. In Section \ref{sec:derivation}, we briefly recap the derivation of the overdamped F\"oppl-von K\'arm\'an equations developed in \cite{Steinbock2022} and in Section \ref{sec:SCE}, we apply the self-consistent expansion (SCE) to these equations to determine the dynamic structure factor of the fluctuating sheet. In Section \ref{sec:simulations}, the accuracy of our solution is confirmed by comparison with numerical simulations. Finally, the implications of these results are discussed in Section \ref{sec:discussion}.

\section{The Overdamped Dynamic F\"oppl-von K\'arm\'an Equations}
\label{sec:derivation}

In \cite{Steinbock2022}, we developed a model to describe fluctuating elastic thin sheets. In this section, we recap the main ideas and relate them to the dynamic structure factor of such a system.

The equilibrium out-of-plane displacement $\xi\left(x,y\right)$ of a thin elastic sheet subject to an external pressure $P_\textrm{ex}$ is given by the well known F\"oppl-von K\'arm\'an equations~\cite{LandauLifshitzElasticityBook}
\begin{align}
P_\textrm{ex} &= \frac{Eh^{3}}{12\left(1-\nu^{2}\right)}\nabla^{4}\xi \nonumber\\
&-h\left(\frac{\partial^2\xi}{\partial x^{2}}\frac{\partial^{2}\chi}{\partial y^{2}}+\frac{\partial^{2}\xi}{\partial y^{2}}\frac{\partial^{2}\chi}{\partial x^{2}}-2\frac{\partial^{2}\xi}{\partial x\partial y}\frac{\partial^{2}\chi}{\partial x\partial y}\right)\\
0 &= \nabla^{4}\chi+E\left[\frac{\partial^{2}\xi}{\partial x^{2}}\frac{\partial^{2}\xi}{\partial y^{2}}-\left(\frac{\partial^{2}\xi}{\partial x\partial y}\right)^{2}\right] \, , \label{eq:fvk airy}
\end{align}
where $h$, $E$ and $\nu$ denote the sheet thickness, Young's modulus and Poisson ratio respectively. The scalar field $\chi\left(x,y\right)$ denotes the Airy stress potential of the deformation. By writing the out-of-plane displacement $\xi\left(x,y\right)$ in the Monge parameterisation, ie. as a function of $x$ and $y$, we assume that deformations of our sheet are mostly flat and thus our focus here will be on weak fluctuations.

To explore the dynamics of a driven fluctuating sheet, we apply Newton's second law to each element of the sheet with density $\rho$
\begin{equation}
h\rho\frac{\partial^{2}\xi}{\partial t^{2}}=-P_\textrm{ex}+P_\textrm{damping}+P_\textrm{driving}\,.
\end{equation}
where $P_\textrm{driving}$ and $P_\textrm{damping}$ describe driving and damping forces respectively. Though variations of this equation have been studied in the context of wave turbulence \cite{During2006, Boudaoud2008, Mordant2008, Cadot2008, Cobelli2009, Humbert2013, Miquel2013, During2015, During2017, During2019, Hassaini2019}, here we continue the approach introduced in \cite{Steinbock2022} of a sheet subject to ordinary fluid friction $P_\textrm{damping}=-\alpha\frac{\partial\xi}{\partial t}$ being driven by conserved Gaussian noise $P_\textrm{driving}=\eta\left(\vec{r},t\right)$ with noise amplitude $D$, that is,
\begin{align}
\left\langle \eta\left(\vec{r},t\right)\right\rangle &=0\\
\left\langle \eta\left(\vec{r},t\right)\eta\left(\vec{r}\,',t'\right)\right\rangle &=-D\delta\left(t-t'\right)\nabla^{2}\delta\left(\vec{r}-\vec{r}\,'\right)\,.
\end{align}
Other driving forces could be considered, but as argued in \cite{Steinbock2022}, there is value in studying the setup where the sheet's center of mass does not wander in space and hence we impose conserved noise on the sheet. More specific forms of noise which are consistent with the conservation of center of mass could also be considered but following the principle of parsimony, we consider only the simplest possibility here.

Taking the overdamped limit and thus neglecting the inertia term $h\rho\frac{\partial^{2}\xi}{\partial t^{2}}$, this approach provides a concrete model for a driven fluctuating elastic sheet, namely, the overdamped dynamic F\"oppl-von K\'arm\'an equation
\begin{multline}
\alpha\frac{\partial\xi}{\partial t}+\frac{Eh^{3}}{12\left(1-\nu^{2}\right)}\nabla^{4}\xi\\
-h\left(\frac{\partial^2\xi}{\partial x^{2}}\frac{\partial^{2}\chi}{\partial y^{2}}+\frac{\partial^{2}\xi}{\partial y^{2}}\frac{\partial^{2}\chi}{\partial x^{2}}-2\frac{\partial^{2}\xi}{\partial x\partial y}\frac{\partial^{2}\chi}{\partial x\partial y}\right)=\eta\left(\vec{r},t\right)\,.\label{eq:fvk dynamic}
\end{multline}
where the Airy stress potential $\chi\left(x,y\right)$ is still determined by equation (\ref{eq:fvk airy}).

The fundamental difference between the problem under study here and the one studied in the wave turbulence community \cite{During2006, Boudaoud2008, Mordant2008, Cadot2008, Cobelli2009, Humbert2013, Miquel2013, During2015, During2017, During2019, Hassaini2019} is that they focus on the regime where inertia is very important, while friction is present only at the smallest scales. Also, the forcing of the sheet, which is often modeled as white noise, is applied only at the largest scales. As a result, the main feature which is studied is the energy cascade from the large scales (where the forcing is applied) to the smallest scales (where it is dissipated). In fact, there exist concrete predictions regarding this energy cascade depending on the specific scenario that drives this cascade. In contrast, we focus on the dynamics of the structure of the sheet under forcing and friction across all scales.

It is shown in \cite{Steinbock2022} that for a sheet with dimensions $L\times L$, equations (\ref{eq:fvk airy}) and (\ref{eq:fvk dynamic}) can be combined into a single equation for the Fourier components $\tilde{\xi}_{\vec{n}}\left(t\right)$ of $\xi\left(x,y,t\right)=\sum_{\vec{n}}\tilde{\xi}_{\vec{n}}\left(t\right)e^{i\frac{2\pi}{L}\vec{n}\cdot\vec{r}}$ where the sum is taken over all lattice points of $\mathbb{Z}^{2}$. After nondimensionalising, one obtains the following Langevin equation 
\begin{multline}
\frac{\partial\bar{\xi}_{\vec{n}}}{\partial\bar{t}}+g\left|\vec{n}\right|^{4}\bar{\xi}_{\vec{n}}\\
+\frac{1}{2}\sum_{\vec{\ell}_{1}\ne\vec{n}}\sum_{\vec{\ell}_{2}}\sum_{\vec{\ell}_{3}}V_{\vec{n},\vec{\ell}_{1},\vec{\ell}_{2},\vec{\ell}_{3}}\bar{\xi}_{\vec{\ell}_{1}}\bar{\xi}_{\vec{\ell}_{2}}\bar{\xi}_{\vec{\ell}_{3}}=\bar{\eta}_{\vec{n}}\left(\bar{t}\,\right)\label{eq:fvk fourier dmnless}
\end{multline}
containing a single dimensionless parameter 
\begin{equation}
g=\frac{2\pi}{12\left(1-\nu^{2}\right)}\sqrt{\frac{\alpha h^{5}E}{D}}\,.
\end{equation}
The scaled time and Fourier components are given by
\begin{align}
\bar{t}&=\left[\left(2\pi\right)^{6}hDE/\left(\alpha^{3}L^{8}\right)\right]^{1/2}t\\
\bar{\xi}_{\vec{n}} &=\left[\left(2\pi\right)^{2}\alpha hE/D\right]^{1/4}\tilde{\xi}_{\vec{n}}
\end{align}
and the dimensionless noise in Fourier space has mean 0 and variance
\begin{equation}
\left\langle \bar{\eta}_{\vec{n}}\left(\bar{t}\,\right)\bar{\eta}_{\vec{n}'}\left(\bar{t}\,'\right)\right\rangle =\left|\vec{n}\right|^{2}\delta_{\vec{n},-\vec{n}'}\delta\left(\bar{t}-\bar{t}\,'\right)\,.
\end{equation}
Finally, the kernel $V_{\vec{n},\vec{\ell}_{1},\vec{\ell}_{2},\vec{\ell}_{3}}$ is simply the Fourier transform of the transverse projection operator of the sheet deformation \cite{NelsonBook2004, Nelson1987} and is given by
\begin{equation}
V_{\vec{n},\vec{\ell}_{1},\vec{\ell}_{2},\vec{\ell}_{3}}=\delta_{\vec{n},\vec{\ell}_{1}+\vec{\ell}_{2}+\vec{\ell}_{3}}\frac{\left|\vec{n}\times\vec{\ell}_{1}\right|^{2}\left|\vec{\ell}_{2}\times\vec{\ell}_{3}\right|^{2}}{\left|\vec{n}-\vec{\ell}_{1}\right|^{4}}\,,
\end{equation}
where we have denoted $\left|\vec{n}\times\vec{\ell}\,\right|=n_{x}\ell_{y}-n_{y}\ell_{x}$, thus equation (\ref{eq:fvk fourier dmnless}) can be thought of as a type of $\phi^{4}$-field Langevin equation with a non-trivial spatially varying kernel $V_{\vec{n},\vec{\ell}_{1},\vec{\ell}_{2},\vec{\ell}_{3}}$ \cite{Kleinert2001}. Accordingly, in principle, equation (\ref{eq:fvk fourier dmnless}) can be used to find structure factors such as the time-dependent two-point function
\begin{equation}
S_{\vec{n}}\left(\bar{t},\bar{t}\,'\right)=\left\langle \bar{\xi}_{\vec{n}}\left(\bar{t}\,\right)\bar{\xi}_{-\vec{n}}\left(\bar{t}\,'\right)\right\rangle 
\label{eq:S(t)}
\end{equation}
which in steady-state will only depend on the difference $\Delta\bar{t}=\left|\bar{t}-\bar{t}\,'\right|$ and thus can be written as a function of a single argument as 
\begin{equation}
S_{\vec{n}}\left(\bar{t}\,\right)=\left\langle \bar{\xi}_{\vec{n}}\left(0\right)\bar{\xi}_{-\vec{n}}\left(\bar{t}\,\right)\right\rangle \,.
\end{equation}
Unfortunately, the single dimensionless parameter $g$ in equation (\ref{eq:fvk fourier dmnless}) is coupled to its linear part and in \cite{Steinbock2022}, it is argued that $g$ is typically small since $g\sim0.1$ for a typical sheet of aluminum or steel. Indeed, the scaling $g\sim h^{5/2}$ ensures that for any sufficiently thin sheet, $g$ will be small and thus any expansion around the linear part of equation (\ref{eq:fvk fourier dmnless}) which treats the nonlinear part as a mere correction is guaranteed to fail. Instead, following the success of \cite{Steinbock2022}, we will analyse equation (\ref{eq:fvk fourier dmnless}) by application of the self-consistent expansion (SCE).

\section{The Self-Consistent Expansion}
\label{sec:SCE}

As described in \cite{Steinbock2022}, the self-consistent expansion (SCE) can be thought of as a renormalised perturbation theory \cite{McComb2003} capable of providing series approximations even in the presence of strong coupling. The method has found previous application to the KPZ equation and its variations \cite{Schwartz1992, Schwartz1998, Katzav1999, Katzav2002, Schwartz2002, Katzav2002a, Katzav2003,Katzav2003b, Katzav2004, Katzav2004a}, fracture and wetting fronts \cite{Katzav2006, Katzav2007} and turbulence \cite{Edwards2002}. More relevant to our system, the SCE provides an extremely successful solution to the zero-dimensional $\phi^{4}$-theory giving good results at low orders and exact convergence at high orders \cite{Schwartz2008, Remez2018}. Accordingly, the success of the SCE in determining the static structure factor of a fluctuating sheet in \cite{Steinbock2022} was not entirely unexpected and since the SCE has a natural extension to dynamic quantities, we extend the approach taken in \cite{Steinbock2022} here.

\subsection{The  Fokker-Planck Equation and the SCE}

As in \cite{Steinbock2022}, we begin by writing the Fokker-Planck equation corresponding to equation (\ref{eq:fvk fourier dmnless}) \cite{RiskenBook}
\begin{multline}
\frac{\partial P}{\partial\bar{t}}=\frac{1}{2}\sum_{\vec{n}}\left|\vec{n}\right|^{2}\frac{\partial^{2}P}{\partial\bar{\xi}_{\vec{n}}\partial\bar{\xi}_{-\vec{n}}}+g\sum_{\vec{n}}\left|\vec{n}\right|^{4}\frac{\partial}{\partial\bar{\xi}_{\vec{n}}}\left(\bar{\xi}_{\vec{n}}P\right)
\\+\frac{1}{2}\sum_{\vec{n}}\frac{\partial}{\partial\bar{\xi}_{\vec{n}}}\left[P\sum_{\vec{\ell}_{1}\ne\vec{n}}\sum_{\vec{\ell}_{2}}\sum_{\vec{\ell}_{3}}V_{\vec{n},\vec{\ell}_{1},\vec{\ell}_{2},\vec{\ell}_{3}}\bar{\xi}_{\vec{\ell}_{1}}\bar{\xi}_{\vec{\ell}_{2}}\bar{\xi}_{\vec{\ell}_{3}}\right]\,,
\end{multline}
where $P=P\left(\left\{ \bar{\xi}_{\vec{n}}\left(\bar{t}\,\right)\right\} ,\bar{t}\,\right)$ denotes the probability functional that the system will have a specific configuration, as prescribed by the Fourier components $\left\{ \bar{\xi}_{\vec{n}}\left(\bar{t}\,\right)\right\}$ at time $\bar{t}$. We can multiply this equation by a function of the Fourier components $\mathbb{F}\left(\left\{ \bar{\xi}_{\vec{n}}\left(\bar{t}\,\right)\right\} \right)$ and integrate over all $\bar{\xi}_{\vec{n}}\left(\bar{t}\,\right)$ to obtain the following equation for the expectations
\begin{multline}
\frac{\partial\left\langle \mathbb{F}\right\rangle }{\partial\bar{t}}=\frac{1}{2}\sum_{\vec{n}}\left|\vec{n}\right|^{2}\left\langle \frac{\partial^{2}\mathbb{F}}{\partial\bar{\xi}_{\vec{n}}\partial\bar{\xi}_{-\vec{n}}}\right\rangle -g\sum_{\vec{n}}\left|\vec{n}\right|^{4}\left\langle \frac{\partial\mathbb{F}}{\partial\bar{\xi}_{\vec{n}}}\bar{\xi}_{\vec{n}}\right\rangle
\\-\frac{1}{2}\sum_{\vec{n}}\sum_{\vec{\ell}_{1}\ne\vec{n}}\sum_{\vec{\ell}_{2}}\sum_{\vec{\ell}_{3}}V_{\vec{n},\vec{\ell}_{1},\vec{\ell}_{2},\vec{\ell}_{3}}\left\langle \frac{\partial\mathbb{F}}{\partial\bar{\xi}_{\vec{n}}}\bar{\xi}_{\vec{\ell}_{1}}\bar{\xi}_{\vec{\ell}_{2}}\bar{\xi}_{\vec{\ell}_{3}}\right\rangle
\label{eq:FP moments}
\end{multline}
where we have defined the expectation values
\begin{equation}
\left\langle \mathbb{F}\right\rangle =\int\prod_{\vec{n}}d\bar{\xi}_{\vec{n}}\,\mathbb{F}\left(\left\{ \bar{\xi}_{\vec{n}}\right\} \right)P\left(\left\{ \bar{\xi}_{\vec{n}}\right\} ,\bar{t}\,\right)\,.
\end{equation}

Equation (\ref{eq:FP moments}) can be used to obtain relationships between various moments. For instance, in \cite{Steinbock2022}, it was shown that subbing in $\mathbb{F} = \bar{\xi}_{\vec{n}}\bar{\xi}_{\vec{n}'}$ results in an equation relating the static two-point function $\left\langle \bar{\xi}_{\vec{n}}\bar{\xi}_{\vec{n}'}\right\rangle$ to the static four-point function $\left\langle \bar{\xi}_{\vec{n}_{1}}\bar{\xi}_{\vec{n}_{2}}\bar{\xi}_{\vec{n}_{3}}\bar{\xi}_{\vec{n}_{4}}\right\rangle$. Similarly, to obtain relations for dynamic quantities, we can sub in time dependent functions such as $\mathbb{F}\left(\left\{ \bar{\xi}_{\vec{n}}\left(\bar{t}\,\right)\right\} \right)=\bar{\xi}_{\vec{n}}\left(0\right)\bar{\xi}_{\vec{n}'}\left(\bar{t}\,\right)$ which results in the ODE
\begin{multline}
\frac{\partial\left\langle \bar{\xi}_{\vec{n}}\left(0\right)\bar{\xi}_{\vec{n}'}\left(\bar{t}\,\right)\right\rangle }{\partial\bar{t}}=-g\left|\vec{n}'\right|^{4}\left\langle \bar{\xi}_{\vec{n}}\left(0\right)\bar{\xi}_{\vec{n}'}\left(\bar{t}\,\right)\right\rangle \\
-\frac{1}{2}\sum_{\vec{\ell}_{1}\ne\vec{n}'}\sum_{\vec{\ell}_{2}}\sum_{\vec{\ell}_{3}}V_{\vec{n}',\vec{\ell}_{1},\vec{\ell}_{2},\vec{\ell}_{3}}\left\langle \bar{\xi}_{\vec{n}}\left(0\right)\bar{\xi}_{\vec{\ell}_{1}}\left(\bar{t}\,\right)\bar{\xi}_{\vec{\ell}_{2}}\left(\bar{t}\,\right)\bar{\xi}_{\vec{\ell}_{3}}\left(\bar{t}\,\right)\right\rangle \,.
\label{eq:ODE no SCE}
\end{multline}

This first order non-homogeneous ODE provides the time-dependent two-point function $\left\langle \bar{\xi}_{\vec{n}}\left(0\right)\bar{\xi}_{\vec{n}'}\left(\bar{t}\,\right)\right\rangle$ if given the time-dependent four-point function $\left\langle \bar{\xi}_{\vec{n}}\left(0\right)\bar{\xi}_{\vec{\ell}_{1}}\left(\bar{t}\,\right)\bar{\xi}_{\vec{\ell}_{2}}\left(\bar{t}\,\right)\bar{\xi}_{\vec{\ell}_{3}}\left(\bar{t}\,\right)\right\rangle$. An ODE for the four-point function can of course be obtained by subbing $\mathbb{F}\left(\left\{ \bar{\xi}_{\vec{n}}\left(\bar{t}\,\right)\right\} \right)=\bar{\xi}_{\vec{n}}\left(0\right)\bar{\xi}_{\vec{\ell}_{1}}\left(\bar{t}\,\right)\bar{\xi}_{\vec{\ell}_{2}}\left(\bar{t}\,\right)\bar{\xi}_{\vec{\ell}_{3}}\left(\bar{t}\,\right)$ into equation (\ref{eq:FP moments}) though the resulting ODE would require knowledge of the time-dependent six-point function. As observed in \cite{Steinbock2022}, this situation of needing higher moments to find lower ones is similar to the BBGKY hierarchy \cite{Balescu1975Book, PlischkeBergersen1994Book, Kardar2007Book} and finding closure is in general non-trivial. The naive approach would be to simply neglect the non-homogeneous part of equation~(\ref{eq:ODE no SCE}) and then attempt to perturbatively correct for it however since the small parameter $g$ is coupled to the homogeneous part of equation~(\ref{eq:ODE no SCE}), the non-homogeneous contribution is large and non-negligible and thus such an approach is guaranteed to fail. Since this occurs at every level of the hierarchy, a more sophisticated approach is required. 

Following the approach taken in \cite{Steinbock2022}, we apply the SCE to equation~(\ref{eq:FP moments}) by introducing a free parameter $\Gamma_{\left|\vec{n}\right|}$
\begin{multline}
\frac{\partial\left\langle \mathbb{F}\right\rangle }{\partial\bar{t}}=\frac{1}{2}\sum_{\vec{n}}\left|\vec{n}\right|^{2}\left\langle \frac{\partial^{2}\mathbb{F}}{\partial\bar{\xi}_{\vec{n}}\partial\bar{\xi}_{-\vec{n}}}\right\rangle-\sum_{\vec{n}}\Gamma_{\left|\vec{n}\right|}\left\langle \frac{\partial\mathbb{F}}{\partial\bar{\xi}_{\vec{n}}}\bar{\xi}_{\vec{n}}\right\rangle\\ 
-\sum_{\vec{n}}\left(g\left|\vec{n}\right|^{4}-\Gamma_{\left|\vec{n}\right|}\right)\left\langle \frac{\partial\mathbb{F}}{\partial\bar{\xi}_{\vec{n}}}\bar{\xi}_{\vec{n}}\right\rangle \\
-\frac{1}{2}\sum_{\vec{n}}\sum_{\vec{\ell}_{1}\ne\vec{n}}\sum_{\vec{\ell}_{2}}\sum_{\vec{\ell}_{3}}V_{\vec{n},\vec{\ell}_{1},\vec{\ell}_{2},\vec{\ell}_{3}}\left\langle \frac{\partial\mathbb{F}}{\partial\bar{\xi}_{\vec{n}}}\bar{\xi}_{\vec{\ell}_{1}}\bar{\xi}_{\vec{\ell}_{2}}\bar{\xi}_{\vec{\ell}_{3}}\right\rangle\,.
\end{multline}
One can think of $\Gamma_{\left|\vec{n}\right|}$ as an effective coupling constant such that a perturbative expansion around the linear theory with $\Gamma_{\left|\vec{n}\right|}$ is valid. The problem of determining its value will be deferred to later though due to the isotropic character of our system, we have already assumed that $\Gamma_{\left|\vec{n}\right|}$ can only depend on the magnitude of $\vec{n}$ and not its direction. Now if $\left\langle \mathbb{F}\right\rangle ^{\left(m\right)}$ denotes an $m^{th}$ order expansion of $\left\langle \mathbb{F}\right\rangle$, then by assumption, the latter terms will contribute at a higher order and thus we can write the iterative relation
\begin{multline}
\frac{\partial\left\langle \mathbb{F}\right\rangle ^{\left(m\right)}}{\partial\bar{t}}=\frac{1}{2}\sum_{\vec{n}}\left|\vec{n}\right|^{2}\left\langle \frac{\partial^{2}\mathbb{F}}{\partial\bar{\xi}_{\vec{n}}\partial\bar{\xi}_{-\vec{n}}}\right\rangle ^{\left(m\right)}\\
-\sum_{\vec{n}}\Gamma_{\left|\vec{n}\right|}\left\langle \frac{\partial\mathbb{F}}{\partial\bar{\xi}_{\vec{n}}}\bar{\xi}_{\vec{n}}\right\rangle ^{\left(m\right)} -\sum_{\vec{n}}\left(g\left|\vec{n}\right|^{4}-\Gamma_{\left|\vec{n}\right|}\right)\left\langle \frac{\partial\mathbb{F}}{\partial\bar{\xi}_{\vec{n}}}\bar{\xi}_{\vec{n}}\right\rangle ^{\left(m-1\right)}\\
-\frac{1}{2}\sum_{\vec{n}}\sum_{\vec{\ell}_{1}\ne\vec{n}}\sum_{\vec{\ell}_{2}}\sum_{\vec{\ell}_{3}}V_{\vec{n},\vec{\ell}_{1},\vec{\ell}_{2},\vec{\ell}_{3}}\left\langle \frac{\partial\mathbb{F}}{\partial\bar{\xi}_{\vec{n}}}\bar{\xi}_{\vec{\ell}_{1}}\bar{\xi}_{\vec{\ell}_{2}}\bar{\xi}_{\vec{\ell}_{3}}\right\rangle ^{\left(m-1\right)}\,,
\label{eq:FP SCE}
\end{multline}
supplemented with the convention that for $m=0$ the $(m-1)$ terms drop out.

This equation can now be used to obtain any moment up to any order. For instance, in \cite{Steinbock2022}, it was shown that subbing in $\mathbb{F} = \bar{\xi}_{\vec{n}}\bar{\xi}_{\vec{n}'}$ or $\mathbb{F}=\bar{\xi}_{\vec{n}}\bar{\xi}_{\vec{\ell}_{1}}\bar{\xi}_{\vec{\ell}_{2}}\bar{\xi}_{\vec{\ell}_{3}}$ together with $m=0$ directly results in zeroth order expressions for the static two-point and four-point functions
\begin{equation}
\left\langle \bar{\xi}_{\vec{n}}\bar{\xi}_{\vec{n}'}\right\rangle ^{\left(0\right)}=\frac{\left|\vec{n}\right|^{2}}{2\Gamma_{\left|\vec{n}\right|}}\delta_{\vec{n},-\vec{n}'}
\label{eq:static 2 pt}
\end{equation}
and
\begin{multline}
\left\langle \bar{\xi}_{\vec{n}'}\bar{\xi}_{\vec{\ell}_{1}}\bar{\xi}_{\vec{\ell}_{2}}\bar{\xi}_{\vec{\ell}_{3}}\right\rangle ^{\left(0\right)}=\left\langle \bar{\xi}_{\vec{n}'}\bar{\xi}_{\vec{\ell}_{1}}\right\rangle ^{\left(0\right)}\left\langle \bar{\xi}_{\vec{\ell}_{2}}\bar{\xi}_{\vec{\ell}_{3}}\right\rangle ^{\left(0\right)}\\
+\left\langle \bar{\xi}_{\vec{n}'}\bar{\xi}_{\vec{\ell}_{2}}\right\rangle ^{\left(0\right)}\left\langle \bar{\xi}_{\vec{\ell}_{1}}\bar{\xi}_{\vec{\ell}_{3}}\right\rangle ^{\left(0\right)}+\left\langle \bar{\xi}_{\vec{n}'}\bar{\xi}_{\vec{\ell}_{3}}\right\rangle ^{\left(0\right)}\left\langle \bar{\xi}_{\vec{\ell}_{1}}\bar{\xi}_{\vec{\ell}_{2}}\right\rangle ^{\left(0\right)}\,.
\label{eq:static 4 pt}
\end{multline}

Upon further substitution of $\mathbb{F} = \bar{\xi}_{\vec{n}}\bar{\xi}_{\vec{n}'}$ and $m=1$, an expression for the first order static two-point function $\left\langle \bar{\xi}_{\vec{n}}\bar{\xi}_{\vec{n}'}\right\rangle ^{\left(1\right)}$ was obtained in terms of the zeroth order static two-point and four-point functions. Here, we will determine what equation~(\ref{eq:FP SCE}) has to say about dynamic time-dependent quantities.

\subsection{Time-Dependent SCE}

Unlike the static quantities described in \cite{Steinbock2022}, subbing time-dependent quantities into equation~(\ref{eq:FP SCE}) does not result in simple expressions for the moments under consideration. Rather, the time derivative on the left-hand side of equation~(\ref{eq:FP SCE}) ensures that time-dependent quantities are given by ODEs. For instance, subbing $\mathbb{F}\left(\left\{ \bar{\xi}_{\vec{n}}\left(\bar{t}\,\right)\right\} \right)=\bar{\xi}_{\vec{n}}\left(0\right)\bar{\xi}_{\vec{n}'}\left(\bar{t}\,\right)$ and $m=0$ into equation~(\ref{eq:FP SCE}) gives the homogeneous ODE
\begin{equation}
\frac{\partial\left\langle \bar{\xi}_{\vec{n}}\left(0\right)\bar{\xi}_{\vec{n}'}\left(\bar{t}\,\right)\right\rangle ^{\left(0\right)}}{\partial\bar{t}}=-\Gamma_{\left|\vec{n}'\right|}\left\langle \bar{\xi}_{\vec{n}}\left(0\right)\bar{\xi}_{\vec{n}'}\left(\bar{t}\,\right)\right\rangle ^{\left(0\right)}
\end{equation}
whose solution is simply
\begin{equation}
\left\langle \bar{\xi}_{\vec{n}}\left(0\right)\bar{\xi}_{\vec{n}'}\left(\bar{t}\,\right)\right\rangle ^{\left(0\right)}=\left\langle \bar{\xi}_{\vec{n}}\bar{\xi}_{\vec{n}'}\right\rangle ^{\left(0\right)}e^{-\Gamma_{\left|\vec{n}'\right|}\bar{t}}
\label{eq:0th order 2pt}
\end{equation}
where $\left\langle \bar{\xi}_{\vec{n}}\bar{\xi}_{\vec{n}'}\right\rangle ^{\left(0\right)}$ is given by equation~(\ref{eq:static 2 pt}).

Similarly, subbing $\mathbb{F}=\bar{\xi}_{\vec{n}'}\left(0\right)\bar{\xi}_{\vec{\ell}_{1}}\left(\bar{t}\,\right)\bar{\xi}_{\vec{\ell}_{2}}\left(\bar{t}\,\right)\bar{\xi}_{\vec{\ell}_{3}}\left(\bar{t}\,\right)$ and $m=0$ into equation (\ref{eq:FP SCE}) gives the ODE
\begin{multline}
\frac{\partial\left\langle \bar{\xi}_{\vec{n}'}\left(0\right)\bar{\xi}_{\vec{\ell}_{1}}\left(\bar{t}\,\right)\bar{\xi}_{\vec{\ell}_{2}}\left(\bar{t}\,\right)\bar{\xi}_{\vec{\ell}_{3}}\left(\bar{t}\,\right)\right\rangle ^{\left(0\right)}}{\partial\bar{t}}=\\
-\sum_{\vec{n}}\Gamma_{\left|\vec{n}\right|}\left\langle \bar{\xi}_{\vec{n}'}\left(0\right)\frac{\partial\left[\bar{\xi}_{\vec{\ell}_{1}}\left(\bar{t}\,\right)\bar{\xi}_{\vec{\ell}_{2}}\left(\bar{t}\,\right)\bar{\xi}_{\vec{\ell}_{3}}\left(\bar{t}\,\right)\right]}{\partial\bar{\xi}_{\vec{n}}\left(\bar{t}\,\right)}\bar{\xi}_{\vec{n}}\left(\bar{t}\,\right)\right\rangle ^{\left(0\right)} \\
+\frac{1}{2}\sum_{\vec{n}}\left|\vec{n}\right|^{2}\left\langle \bar{\xi}_{\vec{n}'}\left(0\right)\frac{\partial^{2}\left[\bar{\xi}_{\vec{\ell}_{1}}\left(\bar{t}\,\right)\bar{\xi}_{\vec{\ell}_{2}}\left(\bar{t}\,\right)\bar{\xi}_{\vec{\ell}_{3}}\left(\bar{t}\,\right)\right]}{\partial\bar{\xi}_{\vec{n}}\left(\bar{t}\,\right)\partial\bar{\xi}_{-\vec{n}}\left(\bar{t}\,\right)}\right\rangle ^{\left(0\right)}\,.
\label{eq:4pt ODE}
\end{multline}
Carrying out the derivatives explicitly and computing the sums, the first term on the right-hand side is simply proportional to the dynamic four-point function
\begin{multline}
\sum_{\vec{n}}\Gamma_{\left|\vec{n}\right|}\left\langle \bar{\xi}_{\vec{n}}\left(0\right)\frac{\partial\left(\bar{\xi}_{\vec{\ell}_{1}}\left(\bar{t}\,\right)\bar{\xi}_{\vec{\ell}_{2}}\left(\bar{t}\,\right)\bar{\xi}_{\vec{\ell}_{3}}\left(\bar{t}\,\right)\right)}{\partial\bar{\xi}_{\vec{n}}\left(\bar{t}\,\right)}\bar{\xi}_{\vec{n}}\left(\bar{t}\,\right)\right\rangle ^{\left(0\right)}=\\
\left(\Gamma_{\left|\vec{\ell}_{1}\right|}+\Gamma_{\left|\vec{\ell}_{2}\right|}+\Gamma_{\left|\vec{\ell}_{3}\right|}\right)\left\langle \bar{\xi}_{\vec{n}}\left(0\right)\bar{\xi}_{\vec{\ell}_{1}}\left(\bar{t}\,\right)\bar{\xi}_{\vec{\ell}_{2}}\left(\bar{t}\,\right)\bar{\xi}_{\vec{\ell}_{3}}\left(\bar{t}\,\right)\right\rangle ^{\left(0\right)}\,,
\label{eq:4pt ODE first sum}
\end{multline}
while the second term is composed of a sum of zeroth order dynamic two-point functions
\begin{multline}
\sum_{\vec{n}}\left|\vec{n}\right|^{2}\left\langle \bar{\xi}_{\vec{n}'}\left(0\right)\frac{\partial^{2}\left(\bar{\xi}_{\vec{\ell}_{1}}\left(\bar{t}\,\right)\bar{\xi}_{\vec{\ell}_{2}}\left(\bar{t}\,\right)\bar{\xi}_{\vec{\ell}_{3}}\left(\bar{t}\,\right)\right)}{\partial\bar{\xi}_{\vec{n}}\left(\bar{t}\,\right)\partial\bar{\xi}_{-\vec{n}}\left(\bar{t}\,\right)}\right\rangle ^{\left(0\right)}=\\
\delta_{\vec{\ell}_{1},-\vec{\ell}_{2}}\left(\left|\vec{\ell}_{1}\right|^{2}+\left|\vec{\ell}_{2}\right|^{2}\right)\left\langle \bar{\xi}_{\vec{n}'}\left(0\right)\bar{\xi}_{\vec{\ell}_{3}}\left(\bar{t}\,\right)\right\rangle ^{\left(0\right)} \\
+\delta_{\vec{\ell}_{1},-\vec{\ell}_{3}}\left(\left|\vec{\ell}_{1}\right|^{2}+\left|\vec{\ell}_{3}\right|^{2}\right)\left\langle \bar{\xi}_{\vec{n}'}\left(0\right)\bar{\xi}_{\vec{\ell}_{2}}\left(\bar{t}\,\right)\right\rangle ^{\left(0\right)} \\
+\delta_{\vec{\ell}_{2},-\vec{\ell}_{3}}\left(\left|\vec{\ell}_{2}\right|^{2}+\left|\vec{\ell}_{3}\right|^{2}\right)\left\langle \bar{\xi}_{\vec{n}'}\left(0\right)\bar{\xi}_{\vec{\ell}_{1}}\left(\bar{t}\,\right)\right\rangle ^{\left(0\right)}\,.
\label{eq:4pt ODE second sum}
\end{multline}
Since we have already computed the zeroth order dynamic two-point function in equation (\ref{eq:0th order 2pt}) and can rearrange equation (\ref{eq:static 2 pt}) to relate $\left|\vec{n}\right|^{2}\delta_{\vec{n},-\vec{n}'}=2\Gamma_{\left|\vec{n}\right|}\left\langle \bar{\xi}_{\vec{n}}\bar{\xi}_{\vec{n}'}\right\rangle ^{\left(0\right)}$, we can observe that
\begin{multline}
\delta_{\vec{\ell}_{1},-\vec{\ell}_{2}}\left(\left|\vec{\ell}_{1}\right|^{2}+\left|\vec{\ell}_{2}\right|^{2}\right)\left\langle \bar{\xi}_{\vec{n}'}\left(0\right)\bar{\xi}_{\vec{\ell}_{3}}\left(\bar{t}\,\right)\right\rangle ^{\left(0\right)}=\\
2\left(\Gamma_{\left|\vec{\ell}_{1}\right|}+\Gamma_{\left|\vec{\ell}_{2}\right|}+\Gamma_{\left|\vec{\ell}_{3}\right|}-\Gamma_{\left|\vec{n}'\right|}\right) \times \\
\times \left\langle \bar{\xi}_{\vec{\ell}_{1}}\bar{\xi}_{\vec{\ell}_{2}}\right\rangle ^{\left(0\right)}\left\langle \bar{\xi}_{\vec{\ell}_{3}}\bar{\xi}_{\vec{n}'}\right\rangle ^{\left(0\right)}e^{-\Gamma_{\left|\vec{n}'\right|}\bar{t}}\,,
\end{multline}
where we have been able to add and subtract $\Gamma_{\left|\vec{\ell}_{3}\right|}$ and $\Gamma_{\left|\vec{n}'\right|}$ since the factor $\left\langle \bar{\xi}_{\vec{\ell}_{3}}\bar{\xi}_{\vec{n}'}\right\rangle ^{\left(0\right)}\propto\delta_{\vec{\ell}_{3},-\vec{n}'}$ ensures that $\Gamma_{\left|\vec{\ell}_{3}\right|}-\Gamma_{\left|\vec{n}'\right|}=0$. Simplifying each term of equation (\ref{eq:4pt ODE second sum}) in this manner, we obtain
\begin{multline}
\sum_{\vec{n}}\left|\vec{n}\right|^{2}\left\langle \bar{\xi}_{\vec{n}'}\left(0\right)\frac{\partial^{2}\left(\bar{\xi}_{\vec{\ell}_{1}}\left(\bar{t}\,\right)\bar{\xi}_{\vec{\ell}_{2}}\left(\bar{t}\,\right)\bar{\xi}_{\vec{\ell}_{3}}\left(\bar{t}\,\right)\right)}{\partial\bar{\xi}_{\vec{n}}\left(\bar{t}\,\right)\partial\bar{\xi}_{-\vec{n}}\left(\bar{t}\,\right)}\right\rangle ^{\left(0\right)}=\\
2\left(\Gamma_{\left|\vec{\ell}_{1}\right|}+\Gamma_{\left|\vec{\ell}_{2}\right|}+\Gamma_{\left|\vec{\ell}_{3}\right|}-\Gamma_{\left|\vec{n}'\right|}\right)\times\\
\times\left\langle \bar{\xi}_{\vec{n}'}\bar{\xi}_{\vec{\ell}_{1}}\bar{\xi}_{\vec{\ell}_{2}}\bar{\xi}_{\vec{\ell}_{3}}\right\rangle ^{\left(0\right)}e^{-\Gamma_{\left|\vec{n}'\right|}\bar{t}}\,,
\label{eq:4pt ODE second sum 2}
\end{multline}
where $\left\langle \bar{\xi}_{\vec{n}'}\bar{\xi}_{\vec{\ell}_{1}}\bar{\xi}_{\vec{\ell}_{2}}\bar{\xi}_{\vec{\ell}_{3}}\right\rangle ^{\left(0\right)}$ is the static four-point function given by equation~(\ref{eq:static 4 pt}).

Finally, substituting equations (\ref{eq:4pt ODE first sum}) and (\ref{eq:4pt ODE second sum 2}) into equation (\ref{eq:4pt ODE}), we obtain the following non-homogeneous ODE for the dynamic four-point function at zeroth order
\begin{multline}
\frac{\partial\left\langle \bar{\xi}_{\vec{n}'}\left(0\right)\bar{\xi}_{\vec{\ell}_{1}}\left(\bar{t}\,\right)\bar{\xi}_{\vec{\ell}_{2}}\left(\bar{t}\,\right)\bar{\xi}_{\vec{\ell}_{3}}\left(\bar{t}\,\right)\right\rangle ^{\left(0\right)}}{\partial\bar{t}}=\\
-\left(\Gamma_{\left|\vec{\ell}_{1}\right|}+\Gamma_{\left|\vec{\ell}_{2}\right|}+\Gamma_{\left|\vec{\ell}_{3}\right|}\right)\left\langle \bar{\xi}_{\vec{n}'}\left(0\right)\bar{\xi}_{\vec{\ell}_{1}}\left(\bar{t}\,\right)\bar{\xi}_{\vec{\ell}_{2}}\left(\bar{t}\,\right)\bar{\xi}_{\vec{\ell}_{3}}\left(\bar{t}\,\right)\right\rangle ^{\left(0\right)}\\
+\left(\Gamma_{\left|\vec{\ell}_{1}\right|}+\Gamma_{\left|\vec{\ell}_{2}\right|}+\Gamma_{\left|\vec{\ell}_{3}\right|}-\Gamma_{\left|\vec{n}'\right|}\right)\left\langle \bar{\xi}_{\vec{n}'}\bar{\xi}_{\vec{\ell}_{1}}\bar{\xi}_{\vec{\ell}_{2}}\bar{\xi}_{\vec{\ell}_{3}}\right\rangle ^{\left(0\right)}e^{-\Gamma_{\left|\vec{n}'\right|}\bar{t}}
\end{multline}
and it is straightforward to check, though perhaps unsurprising, that this is simply solved by
\begin{multline}
\left\langle \bar{\xi}_{\vec{n}'}\left(0\right)\bar{\xi}_{\vec{\ell}_{1}}\left(\bar{t}\,\right)\bar{\xi}_{\vec{\ell}_{2}}\left(\bar{t}\,\right)\bar{\xi}_{\vec{\ell}_{3}}\left(\bar{t}\,\right)\right\rangle ^{\left(0\right)}= \\
\left\langle \bar{\xi}_{\vec{n}'}\bar{\xi}_{\vec{\ell}_{1}}\bar{\xi}_{\vec{\ell}_{2}}\bar{\xi}_{\vec{\ell}_{3}}\right\rangle ^{\left(0\right)}e^{-\Gamma_{\left|\vec{n}'\right|}\bar{t}}\,.
\label{eq:31}
\end{multline}

Now to study the effect of the nonlinearity, we proceed to higher orders. Subbing $\mathbb{F}\left(\left\{ \bar{\xi}_{\vec{n}}\left(\bar{t}\,\right)\right\} \right)=\bar{\xi}_{\vec{n}}\left(0\right)\bar{\xi}_{\vec{n}'}\left(\bar{t}\,\right)$ and $m=1$ into equation (\ref{eq:FP SCE}) gives after some tedious algebra
\begin{multline}
\frac{\partial\left\langle \bar{\xi}_{\vec{n}}\left(0\right)\bar{\xi}_{\vec{n}'}\left(\bar{t}\,\right)\right\rangle ^{\left(1\right)}}{\partial\bar{t}}=-\Gamma_{\left|\vec{n}'\right|}\left\langle \bar{\xi}_{\vec{n}}\left(0\right)\bar{\xi}_{\vec{n}'}\left(\bar{t}\,\right)\right\rangle ^{\left(1\right)}\\
-\left(g\left|\vec{n}'\right|^{4}-\Gamma_{\left|\vec{n}'\right|}\right)\left\langle \bar{\xi}_{\vec{n}}\left(0\right)\bar{\xi}_{\vec{n}'}\left(\bar{t}\,\right)\right\rangle ^{\left(0\right)}\\
-\frac{1}{2}\sum_{\vec{\ell}_{1}\ne\vec{n}'}\sum_{\vec{\ell}_{2}}\sum_{\vec{\ell}_{3}} V_{\vec{n}',\vec{\ell}_{1},\vec{\ell}_{2},\vec{\ell}_{3}}\times \\
\times\left\langle \bar{\xi}_{\vec{n}}\left(0\right)\bar{\xi}_{\vec{\ell}_{1}}\left(\bar{t}\,\right)\bar{\xi}_{\vec{\ell}_{2}}\left(\bar{t}\,\right)\bar{\xi}_{\vec{\ell}_{3}}\left(\bar{t}\,\right)\right\rangle ^{\left(0\right)}\,,
\end{multline}
or after subbing in our expressions for the zeroth order time-dependent two-point and four-point functions from Eqs. (\ref{eq:0th order 2pt}) and (\ref{eq:31}) respectively
\begin{multline}
\frac{\partial\left\langle \bar{\xi}_{\vec{n}}\left(0\right)\bar{\xi}_{\vec{n}'}\left(\bar{t}\,\right)\right\rangle ^{\left(1\right)}}{\partial\bar{t}}=-\Gamma_{\left|\vec{n}'\right|}\left\langle \bar{\xi}_{\vec{n}}\left(0\right)\bar{\xi}_{\vec{n}'}\left(\bar{t}\,\right)\right\rangle ^{\left(1\right)}\\
-\Biggl[\left(g\left|\vec{n}'\right|^{4}-\Gamma_{\left|\vec{n}'\right|}\right)\left\langle \bar{\xi}_{\vec{n}}\bar{\xi}_{\vec{n}'}\right\rangle ^{\left(0\right)}  \\
+\frac{1}{2}\sum_{\vec{\ell}_{1}\ne\vec{n}'}\sum_{\vec{\ell}_{2}}\sum_{\vec{\ell}_{3}}V_{\vec{n}',\vec{\ell}_{1},\vec{\ell}_{2},\vec{\ell}_{3}}\left\langle \bar{\xi}_{\vec{n}}\bar{\xi}_{\vec{\ell}_{1}}\bar{\xi}_{\vec{\ell}_{2}}\bar{\xi}_{\vec{\ell}_{3}}\right\rangle ^{\left(0\right)}\Biggr]e^{-\Gamma_{\left|\vec{n}'\right|}\bar{t}}\,.
\end{multline}

It is worth noting that being precise, the exponential decay associated with the four-point function in this expression should decay with rate $\Gamma_{\left|\vec{n}\right|}$ instead of $\Gamma_{\left|\vec{n}'\right|}$. A careful analysis of the sum over the kernel $V_{\vec{n}',\vec{\ell}_{1},\vec{\ell}_{2},\vec{\ell}_{3}}$ with the static four-point function reveals however that this term vanishes unless $\vec{n}=-\vec{n}'$ and thus no harm is done by replacing $\Gamma_{\left|\vec{n}\right|}$ with $\Gamma_{\left|\vec{n}'\right|}$. It is now apparent that the non-homogeneous term in this equation decays at the natural decay rate $\Gamma_{\left|\vec{n}'\right|}$ of the ODE and thus its general solution will contain a non-physical secular term of the form $\bar{t}e^{-\Gamma_{\left|\vec{n}'\right|}\bar{t}}$. As in the Poincar\'e-Lindstedt method for perturbatively solving non-linear ODEs \cite{Poincare1893, Lindstedt1882, DrazinBook}, this situation can be avoided by setting $\Gamma_{\left|\vec{n}'\right|}$ such that the coefficient of the non-homogeneous term vanishes, ie.
\begin{multline}
0=\left(g\left|\vec{n}'\right|^{4}-\Gamma_{\left|\vec{n}'\right|}\right)\left\langle \bar{\xi}_{\vec{n}}\bar{\xi}_{\vec{n}'}\right\rangle ^{\left(0\right)} \\
+\frac{1}{2}\sum_{\vec{\ell}_{1}\ne\vec{n}'}\sum_{\vec{\ell}_{2}}\sum_{\vec{\ell}_{3}}V_{\vec{n}',\vec{\ell}_{1},\vec{\ell}_{2},\vec{\ell}_{3}}\left\langle \bar{\xi}_{\vec{n}}\bar{\xi}_{\vec{\ell}_{1}}\bar{\xi}_{\vec{\ell}_{2}}\bar{\xi}_{\vec{\ell}_{3}}\right\rangle ^{\left(0\right)}\,.
\end{multline}

\noindent
Use of this idea to determine the characteristic decay rate is different in the context of the SCE method and might be useful in other problems as well.
After subbing in the static two-point and four-point functions and carrying out the sums in the last equation, this ultimately simplifies to the following discrete integral equation for $\Gamma_{\left|\vec{n}\right|}$
\begin{equation}
\Gamma_{\left|\vec{n}\right|}=g\left|\vec{n}\right|^{4}+\frac{1}{2}\sum_{\vec{\ell}\ne\vec{n}}\frac{\left|\vec{\ell}\,\right|^{2}\left|\vec{n}\times\vec{\ell}\,\right|^{4}}{\Gamma_{\left|\vec{\ell}\,\right|}\left|\vec{n}-\vec{\ell}\,\right|^{4}}\,.
\label{eq:disc integ equ}
\end{equation}

Surprisingly, the above argument for preventing secular terms occurring in our expression for $\left\langle \bar{\xi}_{\vec{n}}\left(0\right)\bar{\xi}_{\vec{n}'}\left(\bar{t}\,\right)\right\rangle ^{\left(1\right)}$ is equivalent to simply imposing that the first order approximation for the dynamic structure factor equals its zeroth order approximation, ie.
\begin{equation}
\left\langle \bar{\xi}_{\vec{n}}\left(0\right)\bar{\xi}_{\vec{n}'}\left(\bar{t}\,\right)\right\rangle ^{\left(1\right)}=\left\langle \bar{\xi}_{\vec{n}}\left(0\right)\bar{\xi}_{\vec{n}'}\left(\bar{t}\,\right)\right\rangle ^{\left(0\right)}\,,
\end{equation}
or in other words, we select $\Gamma_{\left|\vec{n}\right|}$ such that the zeroth order approximation is exact up to first order. In \cite{Steinbock2022}, a static version of this self-consistent argument was made by setting the first order approximation for the static structure factor equal to its zeroth order approximation, ie.
\begin{equation}
\left\langle \bar{\xi}_{\vec{n}}\bar{\xi}_{\vec{n}'}\right\rangle ^{\left(1\right)}=\left\langle \bar{\xi}_{\vec{n}}\bar{\xi}_{\vec{n}'}\right\rangle ^{\left(0\right)}\,,
\end{equation}
and indeed, the same discrete integral equation is obtained from both approaches. It is important to appreciate that these two self-consistent arguments giving the same discrete integral equation for $\Gamma_{\left|\vec{n}\right|}$ was by no means anticipated nor trivial and in fact, it is known that this does not occur for the ordinary unadorned $\phi^{4}$-model. In such instances, the fact that the two arguments conflict suggests that the effective decay rate $\Gamma_{\left|\vec{n}\right|}$ also needs to be appropriately expanded in a manner analogous to the Poincar\'e-Lindstedt method \cite{Poincare1893, Lindstedt1882, DrazinBook} if we wish our results to have meaning. Conversely, the situation where the two arguments result in the same discrete integral equation implies some degree of quasi-linearity and is indicative that our approach has self-consistently captured a true feature of the system. 

Equation (\ref{eq:disc integ equ}) has been solved in the appendix of \cite{Steinbock2022}. Here we simply bring the final result
\begin{multline}
\Gamma_{\left|\vec{n}\right|}\simeq n^{4}\sqrt{\frac{3\pi}{4}\left[A\left(g\right)-\ln\left(\frac{n}{n_\textrm{max}}\right)\right]} \times\\
\times \left[1+B\left(g\right)\frac{n}{n_\textrm{max}}\right]\,,
\label{eq:Gamma}
\end{multline}
where $n_\textrm{max}$ is an upper-frequency cutoff which must be imposed on the system and $A\left(g\right)$ and $B\left(g\right)$ are constants which only depend on $g$ and have the following small $g$ expansions
\begin{eqnarray}
A&\simeq &0.137+0.336g+0.243g^{2}+0.112g^3+O\left(g^{4}\right), \label{eq:A theory} \\
B&\simeq &-0.265+0.360g-0.395g^{2}+0.311g^3+O\left(g^{4}\right).\qquad \label{eq:B theory}
\end{eqnarray}
It is worth noting that $B\left(g\right)$ primarily determines the behaviour of $\Gamma_{\left|\vec{n}\right|}$ for large $\vec{n}$, ie. when $\left|\vec{n}\right|\rightarrow n_\textrm{max}$, and can be neglected when $\left|\vec{n}\right|$ is small. Accordingly, for small $\vec{n}$, we have obtained that the decay rate $\Gamma_{\left|\vec{n}\right|}$ grows like a logarithmically corrected power law $\sim n^{4}$.

To summarise, we have found that at first order, the dynamic structure factor is given by
\begin{eqnarray}
\left\langle \bar{\xi}_{\vec{n}}\left(0\right)\bar{\xi}_{\vec{n}'}\left(\bar{t}\,\right)\right\rangle ^{\left(1\right)} &=\left\langle \bar{\xi}_{\vec{n}}\bar{\xi}_{\vec{n}'}\right\rangle ^{\left(0\right)}e^{-\Gamma_{\left|\vec{n}\right|}\bar{t}} \\
&=\delta_{\vec{n},-\vec{n}'}\frac{n^{2}}{2\Gamma_{\left|\vec{n}\right|}}e^{-\Gamma_{\left|\vec{n}\right|}\bar{t}} \label{eq:dyn 2 pt first order}
\end{eqnarray}
where $\Gamma_{\left|\vec{n}\right|}$ is given by equation (\ref{eq:Gamma}) and can be seen to simultaneously play the roles of effective coupling constant and effective decay rate.

\section{Comparison with Simulations}
\label{sec:simulations}

As in \cite{Steinbock2022}, our predictions can be validated by numerical integration of equation (\ref{eq:fvk fourier dmnless}) over a square lattice. Since we investigate here the dynamic properties of the simulation rather than its static properties, the simulations must be run for long enough to obtain precise averages of the time-dependent two-point function $S_{\vec{n}}\left(\bar{t}\,\right)$ and thus they must be run for substantially longer than in \cite{Steinbock2022}. The lattice size imposes a finite maximum frequency $n_\textrm{max}$ but since our theory necessitates the existence of an upper-frequency cut-off, this in itself is fine. More pressingly, for increasing maximum frequency $n_\textrm{max}$, the maximum size of the time step $\delta\bar{t}$ that can be used shrinks if the simulation is to remain stable. This presents a trade-off between the maximum resolution in time vs the maximum resolution in space and since the dynamic structure factor can only be extracted from long simulation runs, the size of $n_\textrm{max}$ is sharply constrained by practical considerations. In practice, our simulations were run with a time-step $\delta\bar{t}=10^{-6}$ over a $41\times41$ lattice corresponding to a maximum frequency $n_\textrm{max}=20\sqrt{2}\approx28.3$, though unlike \cite{Steinbock2022} which only used $10^{6}$ time steps per simulation, these simulations were run for $10^{7}$ time steps each. The consequence of these extended simulations is that an enormous amount of data is generated though simulation states which are close to each other in time are practically indistinguishable except at the very largest modes [an expectation based on Eqs. (\ref{eq:Gamma}) and (\ref{eq:dyn 2 pt first order})]. Since these modes equilibrate far more rapidly than the small modes, they are far less interesting in the current context and thus in the interest of keeping memory resources manageable, simulation data was only saved every $10^3$ time steps. Accordingly, the decay rate of modes which decay faster than $\Delta\bar{t} = 10^3\times\delta\bar{t} = 10^{-3}$ is not presented here. Finally, it is worth noting that as in \cite{Steinbock2022}, efficient calculation of the quartic interaction term of equation (\ref{eq:fvk fourier dmnless}) is non-trivial and as there, was achieved by implementing the pseudo-spectral method described in \cite{During2017} in which the quartic interaction is calculated as the Fourier transform of its real-space counterpart though this imposes periodic boundary conditions on the simulation. To achieve precise results, each simulation was run $10$ times for various values of $0\le g\le 1$ and the results averaged.

\begin{figure}
\includegraphics[width=\columnwidth]{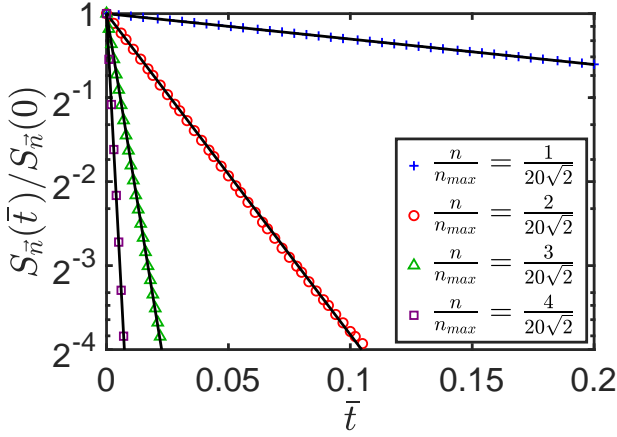}%
\caption{The dynamic structure factor $S_{\vec{n}}\left(\bar{t}\,\right)$ as a function of the time difference $\bar{t}$ for $g=0.1$. Each data set corresponds to a different frequency (see legend) and the solid lines are linear fits to the logarithm of the data. The fact that these fits are excellent confirms that $S_{\vec{n}}\left(\bar{t}\,\right)$ indeed decays exponentially.
\label{fig:decay plot}}
\end{figure}

\begin{figure}
\includegraphics[width=\columnwidth]{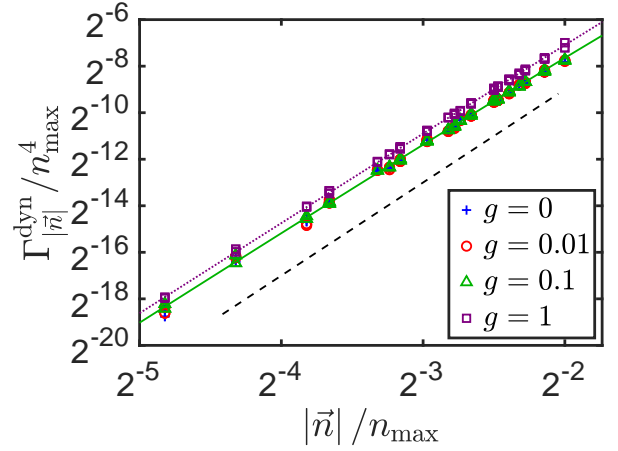}
\includegraphics[width=\columnwidth]{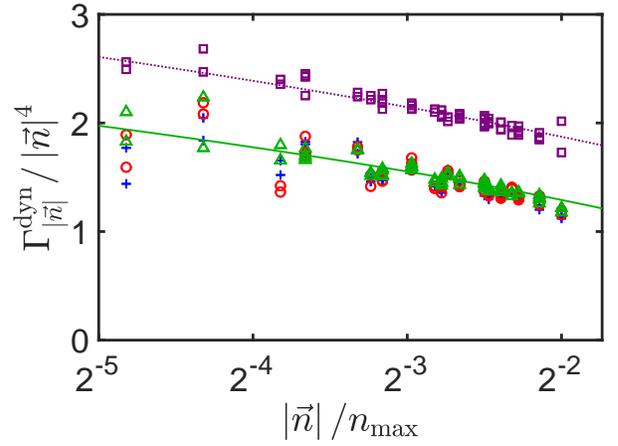}
\caption{$\Gamma_{\left|\vec{n}\right|}^\textrm{dyn}$ (top) and $\Gamma_{\left|\vec{n}\right|}^\textrm{dyn}/\left|\vec{n}\right|^4$ (bottom) as a function of $n$ for various values of $g$. The solid and dotted lines are fits of equation (\ref{eq:fits}) to the $g=0.1$ data (green triangles) and $g=1$ data (purple squares) respectively. The dashed line in the top plot is a guideline proportional to $n^{4}$. The fact that the fits are excellent over the entire range of frequencies confirms that our theory correctly predicts the functional form of the dynamic structure factor $S_{\vec{n}}\left(\bar{t}\,\right)$ and the appreciable deviation between the fits and the guideline shows that the logarithmic correction of equation (\ref{eq:fits}) is non-negligible. This is further emphasised in the bottom plot by the fact that $\Gamma_{\left|\vec{n}\right|}^\textrm{dyn}/\left|\vec{n}\right|^4$ is indeed seen to be non-constant and is well fitted by the theory.
\label{fig:Gamma vs n}}
\end{figure}

Fig. \ref{fig:decay plot} shows how the dynamic structure factor $S_{\vec{n}}\left(\bar{t}\,\right)$ decays for the first few modes as a function of the time difference $\bar{t}$. These results are for the simulation performed with $g=0.1$ though very similar results are obtained for the other values of $g$. The solid black lines are linear fits to the logarithm of the simulation data and clearly show that the dynamic structure factor $S_{\vec{n}}\left(\bar{t}\,\right)$ indeed decays exponentially. By performing such fits for many frequencies, we are able to numerically determine how the decay rate, which we will denote $\Gamma_{\left|\vec{n}\right|}^\textrm{dyn}$, varies as a function of $n$. Fig. \ref{fig:Gamma vs n} shows $\Gamma_{\left|\vec{n}\right|}^\textrm{dyn}$ and $\Gamma_{\left|\vec{n}\right|}^\textrm{dyn}/\left|\vec{n}\right|^4$ as a function of $n$ up to $n_\textrm{max}/4$ for the various values of $g$ we used. The solid and dotted lines are fits of the equation
\begin{equation}
\frac{\Gamma_{\left|\vec{n}\right|}^\textrm{dyn}}{n_\textrm{max}^{4}}\simeq C\left(\frac{n}{n_\textrm{max}}\right)^{4}\sqrt{A\left(g\right)-\ln\left(\frac{n}{n_\textrm{max}}\right)}
\label{eq:fits}
\end{equation}
to the numerical results of $g=0.1$ (green triangles) and $g=1$ (purple squares) respectively. Here, $A\left(g\right)$ was simply taken from equation (\ref{eq:A theory}) while the scaling parameter $C$ was fitted and, as can be seen, these fits are excellent across the entire range of frequencies. As described towards the end of section \ref{sec:SCE}, the constant $B\left(g\right)$ in equation (\ref{eq:Gamma}) primarily modifies $\Gamma_{\left|\vec{n}\right|}$ for large values of $n$ and thus we have not attempted to determine it from our data. Fits for $g=0.01$ (red circles) and $g=0$ (blue pluses) can also be carried out but the resulting fits are so similar to the $g=0.1$ case that little insight is gained by showing them. The dashed line in the top plot of Fig. \ref{fig:Gamma vs n} is a guideline proportional to $n^{4}$ and together with the bottom plot of Fig. \ref{fig:Gamma vs n} clearly emphasises that the logarithmic correction is non-negligible.

\begin{figure}
\includegraphics[width=\columnwidth]{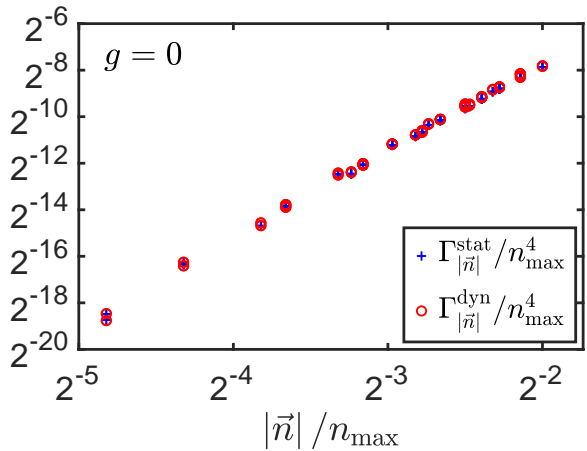}%
\caption{Comparison between the numerical coupling constant $\Gamma_{\left|\vec{n}\right|}^\textrm{stat}=\frac{1}{2}n^{2}/\left\langle \bar{\xi}_{\vec{n}}\bar{\xi}_{-\vec{n}}\right\rangle$ (blue pluses) and the numerical decay rate $\Gamma_{\left|\vec{n}\right|}^\textrm{dyn}$ of the time-dependent two-point function $S_{\vec{n}}\left(\bar{t}\,\right)=S_{\vec{n}}\left(0\right)e^{-\Gamma_{\left|\vec{n}\right|}^\textrm{dyn}\bar{t}}$ (red circles), for $g=0$. These two quantities are seen to agree over all orders of magnitude.
\label{fig:Gamma comparison}}
\end{figure}

The fact that these fits of equation (\ref{eq:fits}) so beautifully capture the behaviour of $\Gamma_{\left|\vec{n}\right|}^\textrm{dyn}$ confirms that our theory has indeed accurately predicted the functional form of the dynamic structure $S_{\vec{n}}\left(\bar{t}\,\right)$. It is worth mentioning that in \cite{Steinbock2022}, it was observed that the periodic square lattice introduces an anisotropy into the simulation and this required each direction to be treated separately, an aspect which we have not observed in the temporal decay data. We explain this distinction by observing that in \cite{Steinbock2022}, the anisotropy was primarily accounted for by scaling the parameter $B\left(g\right)$ in equation (\ref{eq:Gamma}) by some appropriate function $f\left(\theta\right)$ and thus is only observable for large frequencies. Since we do not present here the large frequency decay rate, we have been unable to observe this anisotropy in this data though we presume it too exists.

\begin{figure}
\includegraphics[width=\columnwidth]{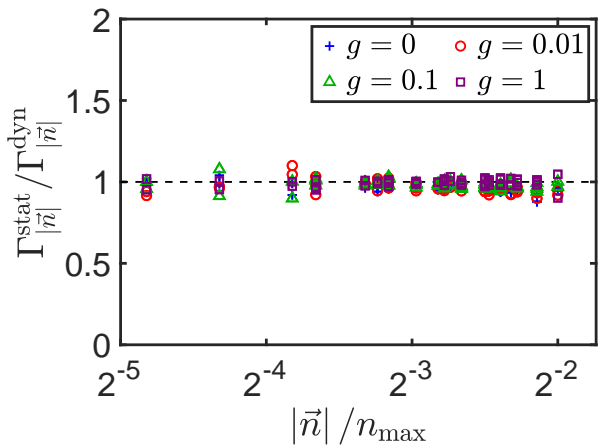}%
\caption{$\Gamma_{\left|\vec{n}\right|}^\textrm{stat}/\Gamma_{\left|\vec{n}\right|}^\textrm{dyn}$ as a function of $n$. For all values of $g$, this ratio is seen to be constant and roughly equal to $1$ over all frequency scales.
\label{fig:Gamma comparison 2}}
\end{figure}

Finally, it is worth comparing $\Gamma_{\left|\vec{n}\right|}^\textrm{dyn}$, the decay rate of the time-dependent structure factor $S_{\vec{n}}\left(\bar{t}\,\right)$, with the effective coupling constant which can be extracted from the relationship
\begin{equation}
S_{\vec{n}}\left(0\right)=\left\langle \bar{\xi}_{\vec{n}}\bar{\xi}_{-\vec{n}}\right\rangle =\frac{n^{2}}{2\Gamma_{\left|\vec{n}\right|}^\textrm{stat}}\,.
\end{equation}
Here, we denote the coupling constant by $\Gamma_{\left|\vec{n}\right|}^\textrm{stat}$ as it is obtained by only considering static quantities. According to the theory developed above and in \cite{Steinbock2022}, $\Gamma_{\left|\vec{n}\right|}^\textrm{stat}$ and $\Gamma_{\left|\vec{n}\right|}^\textrm{dyn}$ should be identical and Fig. \ref{fig:Gamma comparison} and Fig. \ref{fig:Gamma comparison 2} shows that these two methods of obtaining $\Gamma_{\left|\vec{n}\right|}$ are in fact extremely close. Fig. \ref{fig:Gamma comparison} compares $\Gamma_{\left|\vec{n}\right|}^\textrm{stat}$ and $\Gamma_{\left|\vec{n}\right|}^\textrm{dyn}$ for the case of $g=0$ and shows that even with maximal non-linear coupling, $\Gamma_{\left|\vec{n}\right|}^\textrm{stat}$ and $\Gamma_{\left|\vec{n}\right|}^\textrm{dyn}$ are practically indistinguishable. The same result can be obtained for the other values of $g$ and indeed Fig. \ref{fig:Gamma comparison 2} shows that the ratio $\Gamma_{\left|\vec{n}\right|}^\textrm{stat}/\Gamma_{\left|\vec{n}\right|}^\textrm{dyn} \approx 1$ for all values of $g$ over all frequency scales.

\section{Discussion}
\label{sec:discussion}

In this work we have extended a previous paper \cite{Steinbock2022} to study the dynamics of a vibrating thin sheet, governed by the overdamped F\"oppl-von K\'arm\'an equations. Specifically, we applied the self-consistent expansion to obtain predictions which agree with the results of numerical simulations. Surprisingly, the effective coupling constant $\Gamma_{\left|\vec{n}\right|}$, which determines the static structure factor of the sheet coincides with the decay rate of the dynamic structure factor. This observation is reminiscent of linear systems, where these two quantities are indeed the same, yet it is not obvious that the F\"oppl-von K\'arm\'an equations, which are strongly nonlinear, would exhibit such a phenomena. A way to make sense of this situation is by using a recent correlation-response inequality in dynamical systems \cite{Katzav2011a, Katzav2011b}. This inequality was formulated in the context of Langevin dynamics and shows that whenever the equations of motion can be derived from a Hamiltonian and hence obey a certain Fluctuation-Dissipation relation (thus belonging to class I using that classification), and are Galilean invariant (that is, the zero-mode does not affect the dynamics of the higher modes thus belonging to class II in that classification), the exponent inequalities become an equality implying quasi-linear dynamics. The quasi-linearity refers to the fact that the scaling of the decay rate of the dynamic correlation function is related to the static exponents in a simple manner. While the overdamped F\"oppl-von K\'arm\'an equations used here are definitely Galilean invariant since the center of mass does not drift as a result of the dynamics, the way they are derived from a Hamiltonian deviates from the models considered in \cite{Katzav2011a, Katzav2011b}, as they belong to neither model A nor model B in the classical classification of Hohenberg and Halperin \cite{Hohenberg1977}. Nevertheless, they obey the quasi-linear property, which may be the result of a modified Fluctuation-Dissipation relation. This makes the overdamped F\"oppl-von K\'arm\'an equations another interesting example of a physically motivated model that obeys quasi-linearity.

From a methodological point of view, imposing the condition that secular terms should vanish, inspired by the Poincar\'e-Lindstedt method, is a major achievement in the context of the SCE method. In the past this method was only able to yield approximate results for the dynamical structure factor in certain cases, usually based on solving another dedicated approximate integral equation.
In this paper, we gained the insight that the resonant or secular terms that are being generated are actually non-physical, and just like in the Poincar\'e-Linstedt method, should be equated to zero. This insight might become useful in other contexts where the SCE method is applied.
The big surprise here is that, unlike the Poincar\'e-Linstedt method, we do not need to solve a new equation for the dynamic decay rate, since the new equation is identical to the integral equation from the static case. 

The results reported here have experimental implications. Both the structure of thin sheets, as characterised by the static structure factor and the characteristic decay rate could be measured directly. There has already been an attempt to measure the structure of crumpled sheets using its optical signature \cite{Rad2019}, and it is reasonable that with improved imaging techniques leading to faster analysis of dynamic laser speckle patterns \cite{Ge2021}, the full dynamic structure factor could be effectively sampled. The coincidence of the effective coupling constant and the decay rate due to the quasi-linearity should be checked experimentally and if confirmed could be used to enhance the resolution and accuracy of the measurements by cross validating these two independent sources.

Thin sheets also exhibit a very distinctive acoustic footprint, often known as crackling noise \cite{Kramer1996,Houle1996,Sethna2001, Mendes2010}. Clearly, the dynamic structure factor of the vibrating sheet should be correlated with its acoustic emissions however the coupling between the two is known to be nontrivial. Accordingly, a verified analytical form of the dynamic structure factor could provide a solid staring point that may allow progress in that vein.

Last, since the expression for the dynamic structure factor depends on the elastic properties of the sheet and in particular on its elastic moduli, a precise time-resolved measurement of the dynamics of the sheet could provide an indirect measurement of the elastic constants of the material. This might be advantageous when a more direct mechanical test may be destructive or might simply modify these properties, while vibrating the sheet introduces only a mild perturbation.

An interesting extension of this work would be to consider the regime where inertia is important, which is of major interest in the context of wave turbulence in thin sheets
\cite{During2006, Boudaoud2008, Mordant2008, Cadot2008, Cobelli2009, Humbert2013, Miquel2013, During2015, During2017, During2019, Hassaini2019}. In particular, it would be interesting to see whether the growing knowledge on the energy cascade could be qualitatively and quantitatively connected to the structure and dynamics of vibrating sheets. Such a step however would require a major technical advance, namely a nontrivial extension of the self-consistent expansion beyond the overdamped regime, or otherwise the development of an alternative methodology to tackle this problem.

\begin{acknowledgments}
The authors wish to thank Arezki Boudaoud for useful discussions.
This work was supported by the Israel Science Foundation Grant No. 1682/18.
\end{acknowledgments}


\bibliography{bibFile}

\providecommand{\noopsort}[1]{}\providecommand{\singleletter}[1]{#1}%
\begin{thebibliography}{82}%
\makeatletter
\providecommand \@ifxundefined [1]{%
 \@ifx{#1\undefined}
}%
\providecommand \@ifnum [1]{%
 \ifnum #1\expandafter \@firstoftwo
 \else \expandafter \@secondoftwo
 \fi
}%
\providecommand \@ifx [1]{%
 \ifx #1\expandafter \@firstoftwo
 \else \expandafter \@secondoftwo
 \fi
}%
\providecommand \natexlab [1]{#1}%
\providecommand \enquote  [1]{``#1''}%
\providecommand \bibnamefont  [1]{#1}%
\providecommand \bibfnamefont [1]{#1}%
\providecommand \citenamefont [1]{#1}%
\providecommand \href@noop [0]{\@secondoftwo}%
\providecommand \href [0]{\begingroup \@sanitize@url \@href}%
\providecommand \@href[1]{\@@startlink{#1}\@@href}%
\providecommand \@@href[1]{\endgroup#1\@@endlink}%
\providecommand \@sanitize@url [0]{\catcode `\\12\catcode `\$12\catcode
  `\&12\catcode `\#12\catcode `\^12\catcode `\_12\catcode `\%12\relax}%
\providecommand \@@startlink[1]{}%
\providecommand \@@endlink[0]{}%
\providecommand \url  [0]{\begingroup\@sanitize@url \@url }%
\providecommand \@url [1]{\endgroup\@href {#1}{\urlprefix }}%
\providecommand \urlprefix  [0]{URL }%
\providecommand \Eprint [0]{\href }%
\providecommand \doibase [0]{https://doi.org/}%
\providecommand \selectlanguage [0]{\@gobble}%
\providecommand \bibinfo  [0]{\@secondoftwo}%
\providecommand \bibfield  [0]{\@secondoftwo}%
\providecommand \translation [1]{[#1]}%
\providecommand \BibitemOpen [0]{}%
\providecommand \bibitemStop [0]{}%
\providecommand \bibitemNoStop [0]{.\EOS\space}%
\providecommand \EOS [0]{\spacefactor3000\relax}%
\providecommand \BibitemShut  [1]{\csname bibitem#1\endcsname}%
\let\auto@bib@innerbib\@empty
\bibitem [{\citenamefont {Plourabou{\'e}}\ and\ \citenamefont
  {Roux}(1996)}]{PlouraboueRoux1996}%
  \BibitemOpen
  \bibfield  {author} {\bibinfo {author} {\bibfnamefont {F.}~\bibnamefont
  {Plourabou{\'e}}}\ and\ \bibinfo {author} {\bibfnamefont {S.}~\bibnamefont
  {Roux}},\ }\bibfield  {title} {\bibinfo {title} {Experimental study of the
  roughness of crumpled surfaces},\ }\href@noop {} {\bibfield  {journal}
  {\bibinfo  {journal} {Physica A}\ }\textbf {\bibinfo {volume} {227}},\
  \bibinfo {pages} {173} (\bibinfo {year} {1996})}\BibitemShut {NoStop}%
\bibitem [{\citenamefont {Matan}\ \emph {et~al.}(2002)\citenamefont {Matan},
  \citenamefont {Williams}, \citenamefont {Witten},\ and\ \citenamefont
  {Nagel}}]{Matan2002}%
  \BibitemOpen
  \bibfield  {author} {\bibinfo {author} {\bibfnamefont {K.}~\bibnamefont
  {Matan}}, \bibinfo {author} {\bibfnamefont {R.~B.}\ \bibnamefont {Williams}},
  \bibinfo {author} {\bibfnamefont {T.~A.}\ \bibnamefont {Witten}},\ and\
  \bibinfo {author} {\bibfnamefont {S.~R.}\ \bibnamefont {Nagel}},\ }\bibfield
  {title} {\bibinfo {title} {Crumpling a thin sheet},\ }\href
  {https://doi.org/10.1103/PhysRevLett.88.076101} {\bibfield  {journal}
  {\bibinfo  {journal} {Phys. Rev. Lett.}\ }\textbf {\bibinfo {volume} {88}},\
  \bibinfo {pages} {076101} (\bibinfo {year} {2002})}\BibitemShut {NoStop}%
\bibitem [{\citenamefont {Blair}\ and\ \citenamefont
  {Kudrolli}(2005)}]{BlairKudrolli2005}%
  \BibitemOpen
  \bibfield  {author} {\bibinfo {author} {\bibfnamefont {D.~L.}\ \bibnamefont
  {Blair}}\ and\ \bibinfo {author} {\bibfnamefont {A.}~\bibnamefont
  {Kudrolli}},\ }\bibfield  {title} {\bibinfo {title} {Geometry of crumpled
  paper},\ }\href {https://doi.org/10.1103/PhysRevLett.94.166107} {\bibfield
  {journal} {\bibinfo  {journal} {Phys. Rev. Lett.}\ }\textbf {\bibinfo
  {volume} {94}},\ \bibinfo {pages} {166107} (\bibinfo {year}
  {2005})}\BibitemShut {NoStop}%
\bibitem [{\citenamefont {Balankin}\ \emph {et~al.}(2006)\citenamefont
  {Balankin}, \citenamefont {Huerta}, \citenamefont {Cortes Montes~de Oca},
  \citenamefont {Ochoa}, \citenamefont {Mart\'{\i}nez~Trinidad},\ and\
  \citenamefont {Mendoza}}]{Balankin2006}%
  \BibitemOpen
  \bibfield  {author} {\bibinfo {author} {\bibfnamefont {A.~S.}\ \bibnamefont
  {Balankin}}, \bibinfo {author} {\bibfnamefont {O.~S.}\ \bibnamefont
  {Huerta}}, \bibinfo {author} {\bibfnamefont {R.}~\bibnamefont {Cortes
  Montes~de Oca}}, \bibinfo {author} {\bibfnamefont {D.~S.}\ \bibnamefont
  {Ochoa}}, \bibinfo {author} {\bibfnamefont {J.}~\bibnamefont
  {Mart\'{\i}nez~Trinidad}},\ and\ \bibinfo {author} {\bibfnamefont {M.~A.}\
  \bibnamefont {Mendoza}},\ }\bibfield  {title} {\bibinfo {title}
  {Intrinsically anomalous roughness of randomly crumpled thin sheets},\ }\href
  {https://doi.org/10.1103/PhysRevE.74.061602} {\bibfield  {journal} {\bibinfo
  {journal} {Phys. Rev. E}\ }\textbf {\bibinfo {volume} {74}},\ \bibinfo
  {pages} {061602} (\bibinfo {year} {2006})}\BibitemShut {NoStop}%
\bibitem [{\citenamefont {Andresen}\ \emph {et~al.}(2007)\citenamefont
  {Andresen}, \citenamefont {Hansen},\ and\ \citenamefont
  {Schmittbuhl}}]{Andresen2007}%
  \BibitemOpen
  \bibfield  {author} {\bibinfo {author} {\bibfnamefont {C.~A.}\ \bibnamefont
  {Andresen}}, \bibinfo {author} {\bibfnamefont {A.}~\bibnamefont {Hansen}},\
  and\ \bibinfo {author} {\bibfnamefont {J.}~\bibnamefont {Schmittbuhl}},\
  }\bibfield  {title} {\bibinfo {title} {Ridge network in crumpled paper},\
  }\href {https://doi.org/10.1103/PhysRevE.76.026108} {\bibfield  {journal}
  {\bibinfo  {journal} {Phys. Rev. E}\ }\textbf {\bibinfo {volume} {76}},\
  \bibinfo {pages} {026108} (\bibinfo {year} {2007})}\BibitemShut {NoStop}%
\bibitem [{\citenamefont {Balankin}\ and\ \citenamefont
  {Huerta}(2008)}]{Balankin2008}%
  \BibitemOpen
  \bibfield  {author} {\bibinfo {author} {\bibfnamefont {A.~S.}\ \bibnamefont
  {Balankin}}\ and\ \bibinfo {author} {\bibfnamefont {O.~S.}\ \bibnamefont
  {Huerta}},\ }\bibfield  {title} {\bibinfo {title} {Entropic rigidity of a
  crumpling network in a randomly folded thin sheet},\ }\href
  {https://doi.org/10.1103/PhysRevE.77.051124} {\bibfield  {journal} {\bibinfo
  {journal} {Phys. Rev. E}\ }\textbf {\bibinfo {volume} {77}},\ \bibinfo
  {pages} {051124} (\bibinfo {year} {2008})}\BibitemShut {NoStop}%
\bibitem [{\citenamefont {Deboeuf}\ \emph {et~al.}(2013)\citenamefont
  {Deboeuf}, \citenamefont {Katzav}, \citenamefont {Boudaoud}, \citenamefont
  {Bonn},\ and\ \citenamefont {Adda-Bedia}}]{Deboeuf2013}%
  \BibitemOpen
  \bibfield  {author} {\bibinfo {author} {\bibfnamefont {S.}~\bibnamefont
  {Deboeuf}}, \bibinfo {author} {\bibfnamefont {E.}~\bibnamefont {Katzav}},
  \bibinfo {author} {\bibfnamefont {A.}~\bibnamefont {Boudaoud}}, \bibinfo
  {author} {\bibfnamefont {D.}~\bibnamefont {Bonn}},\ and\ \bibinfo {author}
  {\bibfnamefont {M.}~\bibnamefont {Adda-Bedia}},\ }\bibfield  {title}
  {\bibinfo {title} {Comparative study of crumpling and folding of thin
  sheets},\ }\href {https://doi.org/10.1103/PhysRevLett.110.104301} {\bibfield
  {journal} {\bibinfo  {journal} {Phys. Rev. Lett.}\ }\textbf {\bibinfo
  {volume} {110}},\ \bibinfo {pages} {104301} (\bibinfo {year}
  {2013})}\BibitemShut {NoStop}%
\bibitem [{\citenamefont {Balankin}\ \emph {et~al.}(2013)\citenamefont
  {Balankin}, \citenamefont {Horta~Rangel}, \citenamefont
  {Garc\'{\i}a~P\'erez}, \citenamefont {Gayosso~Martinez}, \citenamefont
  {Sanchez~Chavez},\ and\ \citenamefont
  {Mart\'{\i}nez-Gonz\'alez}}]{Balankin2013}%
  \BibitemOpen
  \bibfield  {author} {\bibinfo {author} {\bibfnamefont {A.~S.}\ \bibnamefont
  {Balankin}}, \bibinfo {author} {\bibfnamefont {A.}~\bibnamefont
  {Horta~Rangel}}, \bibinfo {author} {\bibfnamefont {G.}~\bibnamefont
  {Garc\'{\i}a~P\'erez}}, \bibinfo {author} {\bibfnamefont {F.}~\bibnamefont
  {Gayosso~Martinez}}, \bibinfo {author} {\bibfnamefont {H.}~\bibnamefont
  {Sanchez~Chavez}},\ and\ \bibinfo {author} {\bibfnamefont {C.~L.}\
  \bibnamefont {Mart\'{\i}nez-Gonz\'alez}},\ }\bibfield  {title} {\bibinfo
  {title} {Fractal features of a crumpling network in randomly folded thin
  matter and mechanics of sheet crushing},\ }\href
  {https://doi.org/10.1103/PhysRevE.87.052806} {\bibfield  {journal} {\bibinfo
  {journal} {Phys. Rev. E}\ }\textbf {\bibinfo {volume} {87}},\ \bibinfo
  {pages} {052806} (\bibinfo {year} {2013})}\BibitemShut {NoStop}%
\bibitem [{\citenamefont {Lahini}\ \emph {et~al.}(2017)\citenamefont {Lahini},
  \citenamefont {Gottesman}, \citenamefont {Amir},\ and\ \citenamefont
  {Rubinstein}}]{Lahini2017}%
  \BibitemOpen
  \bibfield  {author} {\bibinfo {author} {\bibfnamefont {Y.}~\bibnamefont
  {Lahini}}, \bibinfo {author} {\bibfnamefont {O.}~\bibnamefont {Gottesman}},
  \bibinfo {author} {\bibfnamefont {A.}~\bibnamefont {Amir}},\ and\ \bibinfo
  {author} {\bibfnamefont {S.~M.}\ \bibnamefont {Rubinstein}},\ }\bibfield
  {title} {\bibinfo {title} {Nonmonotonic aging and memory retention in
  disordered mechanical systems},\ }\href
  {https://doi.org/10.1103/PhysRevLett.118.085501} {\bibfield  {journal}
  {\bibinfo  {journal} {Phys. Rev. Lett.}\ }\textbf {\bibinfo {volume} {118}},\
  \bibinfo {pages} {085501} (\bibinfo {year} {2017})}\BibitemShut {NoStop}%
\bibitem [{\citenamefont {Gottesman}\ \emph {et~al.}(2018)\citenamefont
  {Gottesman}, \citenamefont {Andrejevic}, \citenamefont {Rycroft},\ and\
  \citenamefont {Rubinstein}}]{Gottesman2018}%
  \BibitemOpen
  \bibfield  {author} {\bibinfo {author} {\bibfnamefont {O.}~\bibnamefont
  {Gottesman}}, \bibinfo {author} {\bibfnamefont {J.}~\bibnamefont
  {Andrejevic}}, \bibinfo {author} {\bibfnamefont {C.~H.}\ \bibnamefont
  {Rycroft}},\ and\ \bibinfo {author} {\bibfnamefont {S.~M.}\ \bibnamefont
  {Rubinstein}},\ }\bibfield  {title} {\bibinfo {title} {A state variable for
  crumpled thin sheets},\ }\href@noop {} {\bibfield  {journal} {\bibinfo
  {journal} {Communications Physics}\ }\textbf {\bibinfo {volume} {1}},\
  \bibinfo {pages} {1} (\bibinfo {year} {2018})}\BibitemShut {NoStop}%
\bibitem [{\citenamefont {Shohat}\ \emph {et~al.}(2022)\citenamefont {Shohat},
  \citenamefont {Hexner},\ and\ \citenamefont {Lahini}}]{Shohat2022}%
  \BibitemOpen
  \bibfield  {author} {\bibinfo {author} {\bibfnamefont {D.}~\bibnamefont
  {Shohat}}, \bibinfo {author} {\bibfnamefont {D.}~\bibnamefont {Hexner}},\
  and\ \bibinfo {author} {\bibfnamefont {Y.}~\bibnamefont {Lahini}},\
  }\bibfield  {title} {\bibinfo {title} {Memory from coupled instabilities in
  unfolded crumpled sheets},\ }\href {https://doi.org/10.1073/pnas.2200028119}
  {\bibfield  {journal} {\bibinfo  {journal} {Proceedings of the National
  Academy of Sciences}\ }\textbf {\bibinfo {volume} {119}},\ \bibinfo {pages}
  {e2200028119} (\bibinfo {year} {2022})}\BibitemShut {NoStop}%
\bibitem [{\citenamefont {Vliegenthart}\ and\ \citenamefont
  {Gompper}(2006)}]{Vliegenthart2006}%
  \BibitemOpen
  \bibfield  {author} {\bibinfo {author} {\bibfnamefont {G.}~\bibnamefont
  {Vliegenthart}}\ and\ \bibinfo {author} {\bibfnamefont {G.}~\bibnamefont
  {Gompper}},\ }\bibfield  {title} {\bibinfo {title} {Forced crumpling of
  self-avoiding elastic sheets},\ }\href@noop {} {\bibfield  {journal}
  {\bibinfo  {journal} {Nature materials}\ }\textbf {\bibinfo {volume} {5}},\
  \bibinfo {pages} {216} (\bibinfo {year} {2006})}\BibitemShut {NoStop}%
\bibitem [{\citenamefont {Sultan}\ and\ \citenamefont
  {Boudaoud}(2006)}]{Sultan2006}%
  \BibitemOpen
  \bibfield  {author} {\bibinfo {author} {\bibfnamefont {E.}~\bibnamefont
  {Sultan}}\ and\ \bibinfo {author} {\bibfnamefont {A.}~\bibnamefont
  {Boudaoud}},\ }\bibfield  {title} {\bibinfo {title} {Statistics of crumpled
  paper},\ }\href {https://doi.org/10.1103/PhysRevLett.96.136103} {\bibfield
  {journal} {\bibinfo  {journal} {Phys. Rev. Lett.}\ }\textbf {\bibinfo
  {volume} {96}},\ \bibinfo {pages} {136103} (\bibinfo {year}
  {2006})}\BibitemShut {NoStop}%
\bibitem [{\citenamefont {Andrejevic}\ \emph {et~al.}(2021)\citenamefont
  {Andrejevic}, \citenamefont {Lee}, \citenamefont {Rubinstein},\ and\
  \citenamefont {Rycroft}}]{Andrejevic2021}%
  \BibitemOpen
  \bibfield  {author} {\bibinfo {author} {\bibfnamefont {J.}~\bibnamefont
  {Andrejevic}}, \bibinfo {author} {\bibfnamefont {L.~M.}\ \bibnamefont {Lee}},
  \bibinfo {author} {\bibfnamefont {S.~M.}\ \bibnamefont {Rubinstein}},\ and\
  \bibinfo {author} {\bibfnamefont {C.~H.}\ \bibnamefont {Rycroft}},\
  }\bibfield  {title} {\bibinfo {title} {A model for the fragmentation kinetics
  of crumpled thin sheets},\ }\href@noop {} {\bibfield  {journal} {\bibinfo
  {journal} {Nature communications}\ }\textbf {\bibinfo {volume} {12}},\
  \bibinfo {pages} {1470} (\bibinfo {year} {2021})}\BibitemShut {NoStop}%
\bibitem [{\citenamefont {Lobkovsky}\ \emph {et~al.}(1995)\citenamefont
  {Lobkovsky}, \citenamefont {Gentges}, \citenamefont {Li}, \citenamefont
  {Morse},\ and\ \citenamefont {Witten}}]{Lobkovsky1995}%
  \BibitemOpen
  \bibfield  {author} {\bibinfo {author} {\bibfnamefont {A.}~\bibnamefont
  {Lobkovsky}}, \bibinfo {author} {\bibfnamefont {S.}~\bibnamefont {Gentges}},
  \bibinfo {author} {\bibfnamefont {H.}~\bibnamefont {Li}}, \bibinfo {author}
  {\bibfnamefont {D.}~\bibnamefont {Morse}},\ and\ \bibinfo {author}
  {\bibfnamefont {T.~A.}\ \bibnamefont {Witten}},\ }\bibfield  {title}
  {\bibinfo {title} {Scaling properties of stretching ridges in a crumpled
  elastic sheet},\ }\href@noop {} {\bibfield  {journal} {\bibinfo  {journal}
  {Science}\ }\textbf {\bibinfo {volume} {270}},\ \bibinfo {pages} {1482}
  (\bibinfo {year} {1995})}\BibitemShut {NoStop}%
\bibitem [{\citenamefont {Lobkovsky}\ and\ \citenamefont
  {Witten}(1997)}]{Lobkovsky1997}%
  \BibitemOpen
  \bibfield  {author} {\bibinfo {author} {\bibfnamefont {A.~E.}\ \bibnamefont
  {Lobkovsky}}\ and\ \bibinfo {author} {\bibfnamefont {T.~A.}\ \bibnamefont
  {Witten}},\ }\bibfield  {title} {\bibinfo {title} {Properties of ridges in
  elastic membranes},\ }\href {https://doi.org/10.1103/PhysRevE.55.1577}
  {\bibfield  {journal} {\bibinfo  {journal} {Phys. Rev. E}\ }\textbf {\bibinfo
  {volume} {55}},\ \bibinfo {pages} {1577} (\bibinfo {year}
  {1997})}\BibitemShut {NoStop}%
\bibitem [{\citenamefont {Ben~Amar}\ and\ \citenamefont
  {Pomeau}(1997)}]{BenAmar1997}%
  \BibitemOpen
  \bibfield  {author} {\bibinfo {author} {\bibfnamefont {M.}~\bibnamefont
  {Ben~Amar}}\ and\ \bibinfo {author} {\bibfnamefont {Y.}~\bibnamefont
  {Pomeau}},\ }\bibfield  {title} {\bibinfo {title} {Crumpled paper},\
  }\href@noop {} {\bibfield  {journal} {\bibinfo  {journal} {Proceedings of the
  Royal Society of London. Series A: Mathematical, Physical and Engineering
  Sciences}\ }\textbf {\bibinfo {volume} {453}},\ \bibinfo {pages} {729}
  (\bibinfo {year} {1997})}\BibitemShut {NoStop}%
\bibitem [{\citenamefont {Cha\"{\i}eb}\ and\ \citenamefont
  {Melo}(1997)}]{Chaieb1997}%
  \BibitemOpen
  \bibfield  {author} {\bibinfo {author} {\bibfnamefont {S.}~\bibnamefont
  {Cha\"{\i}eb}}\ and\ \bibinfo {author} {\bibfnamefont {F.}~\bibnamefont
  {Melo}},\ }\bibfield  {title} {\bibinfo {title} {Experimental study of crease
  formation in an axially compressed sheet},\ }\href
  {https://doi.org/10.1103/PhysRevE.56.4736} {\bibfield  {journal} {\bibinfo
  {journal} {Phys. Rev. E}\ }\textbf {\bibinfo {volume} {56}},\ \bibinfo
  {pages} {4736} (\bibinfo {year} {1997})}\BibitemShut {NoStop}%
\bibitem [{\citenamefont {Cerda}\ \emph {et~al.}(1999)\citenamefont {Cerda},
  \citenamefont {Chaieb}, \citenamefont {Melo},\ and\ \citenamefont
  {Mahadevan}}]{Cerda1999}%
  \BibitemOpen
  \bibfield  {author} {\bibinfo {author} {\bibfnamefont {E.}~\bibnamefont
  {Cerda}}, \bibinfo {author} {\bibfnamefont {S.}~\bibnamefont {Chaieb}},
  \bibinfo {author} {\bibfnamefont {F.}~\bibnamefont {Melo}},\ and\ \bibinfo
  {author} {\bibfnamefont {L.}~\bibnamefont {Mahadevan}},\ }\bibfield  {title}
  {\bibinfo {title} {Conical dislocations in crumpling},\ }\href@noop {}
  {\bibfield  {journal} {\bibinfo  {journal} {Nature}\ }\textbf {\bibinfo
  {volume} {401}},\ \bibinfo {pages} {46} (\bibinfo {year} {1999})}\BibitemShut
  {NoStop}%
\bibitem [{\citenamefont {Mora}\ and\ \citenamefont
  {Boudaoud}(2002)}]{Mora2002}%
  \BibitemOpen
  \bibfield  {author} {\bibinfo {author} {\bibfnamefont {T.}~\bibnamefont
  {Mora}}\ and\ \bibinfo {author} {\bibfnamefont {A.}~\bibnamefont
  {Boudaoud}},\ }\bibfield  {title} {\bibinfo {title} {Thin elastic plates: On
  the core of developable cones},\ }\href@noop {} {\bibfield  {journal}
  {\bibinfo  {journal} {EPL (Europhysics Letters)}\ }\textbf {\bibinfo {volume}
  {59}},\ \bibinfo {pages} {41} (\bibinfo {year} {2002})}\BibitemShut {NoStop}%
\bibitem [{\citenamefont {Liang}\ and\ \citenamefont
  {Witten}(2005)}]{Liang2005}%
  \BibitemOpen
  \bibfield  {author} {\bibinfo {author} {\bibfnamefont {T.}~\bibnamefont
  {Liang}}\ and\ \bibinfo {author} {\bibfnamefont {T.~A.}\ \bibnamefont
  {Witten}},\ }\bibfield  {title} {\bibinfo {title} {Crescent singularities in
  crumpled sheets},\ }\href {https://doi.org/10.1103/PhysRevE.71.016612}
  {\bibfield  {journal} {\bibinfo  {journal} {Phys. Rev. E}\ }\textbf {\bibinfo
  {volume} {71}},\ \bibinfo {pages} {016612} (\bibinfo {year}
  {2005})}\BibitemShut {NoStop}%
\bibitem [{\citenamefont {Kantor}\ \emph {et~al.}(1986)\citenamefont {Kantor},
  \citenamefont {Kardar},\ and\ \citenamefont {Nelson}}]{Kantor1986}%
  \BibitemOpen
  \bibfield  {author} {\bibinfo {author} {\bibfnamefont {Y.}~\bibnamefont
  {Kantor}}, \bibinfo {author} {\bibfnamefont {M.}~\bibnamefont {Kardar}},\
  and\ \bibinfo {author} {\bibfnamefont {D.~R.}\ \bibnamefont {Nelson}},\
  }\bibfield  {title} {\bibinfo {title} {Statistical mechanics of tethered
  surfaces},\ }\href {https://doi.org/10.1103/PhysRevLett.57.791} {\bibfield
  {journal} {\bibinfo  {journal} {Phys. Rev. Lett.}\ }\textbf {\bibinfo
  {volume} {57}},\ \bibinfo {pages} {791} (\bibinfo {year} {1986})}\BibitemShut
  {NoStop}%
\bibitem [{\citenamefont {Kantor}\ \emph {et~al.}(1987)\citenamefont {Kantor},
  \citenamefont {Kardar},\ and\ \citenamefont {Nelson}}]{Kantor1987}%
  \BibitemOpen
  \bibfield  {author} {\bibinfo {author} {\bibfnamefont {Y.}~\bibnamefont
  {Kantor}}, \bibinfo {author} {\bibfnamefont {M.}~\bibnamefont {Kardar}},\
  and\ \bibinfo {author} {\bibfnamefont {D.~R.}\ \bibnamefont {Nelson}},\
  }\bibfield  {title} {\bibinfo {title} {Tethered surfaces: Statics and
  dynamics},\ }\href {https://doi.org/10.1103/PhysRevA.35.3056} {\bibfield
  {journal} {\bibinfo  {journal} {Phys. Rev. A}\ }\textbf {\bibinfo {volume}
  {35}},\ \bibinfo {pages} {3056} (\bibinfo {year} {1987})}\BibitemShut
  {NoStop}%
\bibitem [{\citenamefont {Frey}\ and\ \citenamefont {Nelson}(1991)}]{Frey1991}%
  \BibitemOpen
  \bibfield  {author} {\bibinfo {author} {\bibfnamefont {E.}~\bibnamefont
  {Frey}}\ and\ \bibinfo {author} {\bibfnamefont {D.~L.}\ \bibnamefont
  {Nelson}},\ }\bibfield  {title} {\bibinfo {title} {{Dynamics of flat
  membranes and flickering in red blood cells}},\ }\href
  {https://doi.org/10.1051/jp1:1991238} {\bibfield  {journal} {\bibinfo
  {journal} {{Journal de Physique I}}\ }\textbf {\bibinfo {volume} {1}},\
  \bibinfo {pages} {1715} (\bibinfo {year} {1991})}\BibitemShut {NoStop}%
\bibitem [{\citenamefont {Niel}(1989)}]{Niel1989}%
  \BibitemOpen
  \bibfield  {author} {\bibinfo {author} {\bibfnamefont {J.}~\bibnamefont
  {Niel}},\ }\bibfield  {title} {\bibinfo {title} {Critical dynamics of
  polymerized membranes at the crumpling transition},\ }\href@noop {}
  {\bibfield  {journal} {\bibinfo  {journal} {EPL (Europhysics Letters)}\
  }\textbf {\bibinfo {volume} {9}},\ \bibinfo {pages} {415} (\bibinfo {year}
  {1989})}\BibitemShut {NoStop}%
\bibitem [{\citenamefont {Steinbock}\ \emph {et~al.}(2022)\citenamefont
  {Steinbock}, \citenamefont {Katzav},\ and\ \citenamefont
  {Boudaoud}}]{Steinbock2022}%
  \BibitemOpen
  \bibfield  {author} {\bibinfo {author} {\bibfnamefont {C.}~\bibnamefont
  {Steinbock}}, \bibinfo {author} {\bibfnamefont {E.}~\bibnamefont {Katzav}},\
  and\ \bibinfo {author} {\bibfnamefont {A.}~\bibnamefont {Boudaoud}},\
  }\bibfield  {title} {\bibinfo {title} {Structure of fluctuating thin sheets
  under random forcing},\ }\href
  {https://doi.org/10.1103/PhysRevResearch.4.033096} {\bibfield  {journal}
  {\bibinfo  {journal} {Phys. Rev. Research}\ }\textbf {\bibinfo {volume}
  {4}},\ \bibinfo {pages} {033096} (\bibinfo {year} {2022})}\BibitemShut
  {NoStop}%
\bibitem [{\citenamefont {Kramer}\ and\ \citenamefont
  {Lobkovsky}(1996)}]{Kramer1996}%
  \BibitemOpen
  \bibfield  {author} {\bibinfo {author} {\bibfnamefont {E.~M.}\ \bibnamefont
  {Kramer}}\ and\ \bibinfo {author} {\bibfnamefont {A.~E.}\ \bibnamefont
  {Lobkovsky}},\ }\bibfield  {title} {\bibinfo {title} {Universal power law in
  the noise from a crumpled elastic sheet},\ }\href
  {https://doi.org/10.1103/PhysRevE.53.1465} {\bibfield  {journal} {\bibinfo
  {journal} {Phys. Rev. E}\ }\textbf {\bibinfo {volume} {53}},\ \bibinfo
  {pages} {1465} (\bibinfo {year} {1996})}\BibitemShut {NoStop}%
\bibitem [{\citenamefont {Houle}\ and\ \citenamefont
  {Sethna}(1996)}]{Houle1996}%
  \BibitemOpen
  \bibfield  {author} {\bibinfo {author} {\bibfnamefont {P.~A.}\ \bibnamefont
  {Houle}}\ and\ \bibinfo {author} {\bibfnamefont {J.~P.}\ \bibnamefont
  {Sethna}},\ }\bibfield  {title} {\bibinfo {title} {Acoustic emission from
  crumpling paper},\ }\href {https://doi.org/10.1103/PhysRevE.54.278}
  {\bibfield  {journal} {\bibinfo  {journal} {Phys. Rev. E}\ }\textbf {\bibinfo
  {volume} {54}},\ \bibinfo {pages} {278} (\bibinfo {year} {1996})}\BibitemShut
  {NoStop}%
\bibitem [{\citenamefont {Sethna}\ \emph {et~al.}(2001)\citenamefont {Sethna},
  \citenamefont {Dahmen},\ and\ \citenamefont {Myers}}]{Sethna2001}%
  \BibitemOpen
  \bibfield  {author} {\bibinfo {author} {\bibfnamefont {J.~P.}\ \bibnamefont
  {Sethna}}, \bibinfo {author} {\bibfnamefont {K.~A.}\ \bibnamefont {Dahmen}},\
  and\ \bibinfo {author} {\bibfnamefont {C.~R.}\ \bibnamefont {Myers}},\
  }\bibfield  {title} {\bibinfo {title} {Crackling noise},\ }\href@noop {}
  {\bibfield  {journal} {\bibinfo  {journal} {Nature}\ }\textbf {\bibinfo
  {volume} {410}},\ \bibinfo {pages} {242} (\bibinfo {year}
  {2001})}\BibitemShut {NoStop}%
\bibitem [{\citenamefont {Mendes}\ \emph {et~al.}(2010)\citenamefont {Mendes},
  \citenamefont {Malacarne}, \citenamefont {Santos}, \citenamefont {Ribeiro},\
  and\ \citenamefont {Picoli}}]{Mendes2010}%
  \BibitemOpen
  \bibfield  {author} {\bibinfo {author} {\bibfnamefont {R.~S.}\ \bibnamefont
  {Mendes}}, \bibinfo {author} {\bibfnamefont {L.~C.}\ \bibnamefont
  {Malacarne}}, \bibinfo {author} {\bibfnamefont {R.~P.~B.}\ \bibnamefont
  {Santos}}, \bibinfo {author} {\bibfnamefont {H.~V.}\ \bibnamefont
  {Ribeiro}},\ and\ \bibinfo {author} {\bibfnamefont {S.}~\bibnamefont
  {Picoli}},\ }\bibfield  {title} {\bibinfo {title} {Earthquake-like patterns
  of acoustic emission in crumpled plastic sheets},\ }\href
  {https://doi.org/10.1209/0295-5075/92/29001} {\bibfield  {journal} {\bibinfo
  {journal} {{EPL} (Europhysics Letters)}\ }\textbf {\bibinfo {volume} {92}},\
  \bibinfo {pages} {29001} (\bibinfo {year} {2010})}\BibitemShut {NoStop}%
\bibitem [{\citenamefont {Rad}\ \emph {et~al.}(2019)\citenamefont {Rad},
  \citenamefont {Ram{\'\i}rez-Miquet}, \citenamefont {Cabrera}, \citenamefont
  {Habibi},\ and\ \citenamefont {Moradi}}]{Rad2019}%
  \BibitemOpen
  \bibfield  {author} {\bibinfo {author} {\bibfnamefont {V.~F.}\ \bibnamefont
  {Rad}}, \bibinfo {author} {\bibfnamefont {E.~E.}\ \bibnamefont
  {Ram{\'\i}rez-Miquet}}, \bibinfo {author} {\bibfnamefont {H.}~\bibnamefont
  {Cabrera}}, \bibinfo {author} {\bibfnamefont {M.}~\bibnamefont {Habibi}},\
  and\ \bibinfo {author} {\bibfnamefont {A.-R.}\ \bibnamefont {Moradi}},\
  }\bibfield  {title} {\bibinfo {title} {Speckle pattern analysis of crumpled
  papers},\ }\href@noop {} {\bibfield  {journal} {\bibinfo  {journal} {Applied
  optics}\ }\textbf {\bibinfo {volume} {58}},\ \bibinfo {pages} {6549}
  (\bibinfo {year} {2019})}\BibitemShut {NoStop}%
\bibitem [{\citenamefont {Mehreganian}\ \emph {et~al.}(2019)\citenamefont
  {Mehreganian}, \citenamefont {Fallah},\ and\ \citenamefont
  {Louca}}]{Mehreganian2019}%
  \BibitemOpen
  \bibfield  {author} {\bibinfo {author} {\bibfnamefont {N.}~\bibnamefont
  {Mehreganian}}, \bibinfo {author} {\bibfnamefont {A.}~\bibnamefont
  {Fallah}},\ and\ \bibinfo {author} {\bibfnamefont {L.}~\bibnamefont
  {Louca}},\ }\bibfield  {title} {\bibinfo {title} {Nonlinear dynamics of
  locally pulse loaded square föppl–von kármán thin plates},\ }\href
  {https://doi.org/https://doi.org/10.1016/j.ijmecsci.2019.105157} {\bibfield
  {journal} {\bibinfo  {journal} {International Journal of Mechanical
  Sciences}\ }\textbf {\bibinfo {volume} {163}},\ \bibinfo {pages} {105157}
  (\bibinfo {year} {2019})}\BibitemShut {NoStop}%
\bibitem [{\citenamefont {Mehreganian}\ \emph {et~al.}(2021)\citenamefont
  {Mehreganian}, \citenamefont {Toolabi}, \citenamefont {Zhuk}, \citenamefont
  {{Etminan Moghadam}}, \citenamefont {Louca},\ and\ \citenamefont
  {Fallah}}]{Mehreganian2021}%
  \BibitemOpen
  \bibfield  {author} {\bibinfo {author} {\bibfnamefont {N.}~\bibnamefont
  {Mehreganian}}, \bibinfo {author} {\bibfnamefont {M.}~\bibnamefont
  {Toolabi}}, \bibinfo {author} {\bibfnamefont {Y.}~\bibnamefont {Zhuk}},
  \bibinfo {author} {\bibfnamefont {F.}~\bibnamefont {{Etminan Moghadam}}},
  \bibinfo {author} {\bibfnamefont {L.}~\bibnamefont {Louca}},\ and\ \bibinfo
  {author} {\bibfnamefont {A.}~\bibnamefont {Fallah}},\ }\bibfield  {title}
  {\bibinfo {title} {Dynamics of pulse-loaded circular föppl-von kármán thin
  plates- analytical and numerical studies},\ }\href
  {https://doi.org/https://doi.org/10.1016/j.jsv.2021.116413} {\bibfield
  {journal} {\bibinfo  {journal} {Journal of Sound and Vibration}\ }\textbf
  {\bibinfo {volume} {513}},\ \bibinfo {pages} {116413} (\bibinfo {year}
  {2021})}\BibitemShut {NoStop}%
\bibitem [{\citenamefont {Nelson}\ \emph {et~al.}(2004)\citenamefont {Nelson},
  \citenamefont {Piran},\ and\ \citenamefont {Weinberg}}]{NelsonBook2004}%
  \BibitemOpen
  \bibfield  {author} {\bibinfo {author} {\bibfnamefont {D.}~\bibnamefont
  {Nelson}}, \bibinfo {author} {\bibfnamefont {T.}~\bibnamefont {Piran}},\ and\
  \bibinfo {author} {\bibfnamefont {S.}~\bibnamefont {Weinberg}},\ }\href@noop
  {} {\emph {\bibinfo {title} {Statistical mechanics of membranes and
  surfaces}}},\ \bibinfo {edition} {2nd}\ ed.\ (\bibinfo  {publisher} {World
  Scientific},\ \bibinfo {address} {Singapore},\ \bibinfo {year}
  {2004})\BibitemShut {NoStop}%
\bibitem [{\citenamefont {Liang}\ and\ \citenamefont
  {Purohit}(2016)}]{Liang2016}%
  \BibitemOpen
  \bibfield  {author} {\bibinfo {author} {\bibfnamefont {X.}~\bibnamefont
  {Liang}}\ and\ \bibinfo {author} {\bibfnamefont {P.~K.}\ \bibnamefont
  {Purohit}},\ }\bibfield  {title} {\bibinfo {title} {A fluctuating elastic
  plate and a cell model for lipid membranes},\ }\href@noop {} {\bibfield
  {journal} {\bibinfo  {journal} {Journal of the Mechanics and Physics of
  Solids}\ }\textbf {\bibinfo {volume} {90}},\ \bibinfo {pages} {29} (\bibinfo
  {year} {2016})}\BibitemShut {NoStop}%
\bibitem [{\citenamefont {Meyer}\ \emph
  {et~al.}(2007{\natexlab{a}})\citenamefont {Meyer}, \citenamefont {Geim},
  \citenamefont {Katsnelson}, \citenamefont {Novoselov}, \citenamefont
  {Booth},\ and\ \citenamefont {Roth}}]{Meyer2007}%
  \BibitemOpen
  \bibfield  {author} {\bibinfo {author} {\bibfnamefont {J.~C.}\ \bibnamefont
  {Meyer}}, \bibinfo {author} {\bibfnamefont {A.~K.}\ \bibnamefont {Geim}},
  \bibinfo {author} {\bibfnamefont {M.~I.}\ \bibnamefont {Katsnelson}},
  \bibinfo {author} {\bibfnamefont {K.~S.}\ \bibnamefont {Novoselov}}, \bibinfo
  {author} {\bibfnamefont {T.~J.}\ \bibnamefont {Booth}},\ and\ \bibinfo
  {author} {\bibfnamefont {S.}~\bibnamefont {Roth}},\ }\bibfield  {title}
  {\bibinfo {title} {The structure of suspended graphene sheets},\ }\href@noop
  {} {\bibfield  {journal} {\bibinfo  {journal} {Nature}\ }\textbf {\bibinfo
  {volume} {446}},\ \bibinfo {pages} {60} (\bibinfo {year}
  {2007}{\natexlab{a}})}\BibitemShut {NoStop}%
\bibitem [{\citenamefont {Meyer}\ \emph
  {et~al.}(2007{\natexlab{b}})\citenamefont {Meyer}, \citenamefont {Geim},
  \citenamefont {Katsnelson}, \citenamefont {Novoselov}, \citenamefont
  {Obergfell}, \citenamefont {Roth}, \citenamefont {Girit},\ and\ \citenamefont
  {Zettl}}]{Meyer2007a}%
  \BibitemOpen
  \bibfield  {author} {\bibinfo {author} {\bibfnamefont {J.~C.}\ \bibnamefont
  {Meyer}}, \bibinfo {author} {\bibfnamefont {A.}~\bibnamefont {Geim}},
  \bibinfo {author} {\bibfnamefont {M.}~\bibnamefont {Katsnelson}}, \bibinfo
  {author} {\bibfnamefont {K.}~\bibnamefont {Novoselov}}, \bibinfo {author}
  {\bibfnamefont {D.}~\bibnamefont {Obergfell}}, \bibinfo {author}
  {\bibfnamefont {S.}~\bibnamefont {Roth}}, \bibinfo {author} {\bibfnamefont
  {C.}~\bibnamefont {Girit}},\ and\ \bibinfo {author} {\bibfnamefont
  {A.}~\bibnamefont {Zettl}},\ }\bibfield  {title} {\bibinfo {title} {On the
  roughness of single-and bi-layer graphene membranes},\ }\href@noop {}
  {\bibfield  {journal} {\bibinfo  {journal} {Solid State Communications}\
  }\textbf {\bibinfo {volume} {143}},\ \bibinfo {pages} {101} (\bibinfo {year}
  {2007}{\natexlab{b}})}\BibitemShut {NoStop}%
\bibitem [{\citenamefont {Fasolino}\ \emph {et~al.}(2007)\citenamefont
  {Fasolino}, \citenamefont {Los},\ and\ \citenamefont
  {Katsnelson}}]{Fasolino2007}%
  \BibitemOpen
  \bibfield  {author} {\bibinfo {author} {\bibfnamefont {A.}~\bibnamefont
  {Fasolino}}, \bibinfo {author} {\bibfnamefont {J.}~\bibnamefont {Los}},\ and\
  \bibinfo {author} {\bibfnamefont {M.~I.}\ \bibnamefont {Katsnelson}},\
  }\bibfield  {title} {\bibinfo {title} {Intrinsic ripples in graphene},\
  }\href@noop {} {\bibfield  {journal} {\bibinfo  {journal} {Nature materials}\
  }\textbf {\bibinfo {volume} {6}},\ \bibinfo {pages} {858} (\bibinfo {year}
  {2007})}\BibitemShut {NoStop}%
\bibitem [{\citenamefont {Thompson-Flagg}\ \emph {et~al.}(2009)\citenamefont
  {Thompson-Flagg}, \citenamefont {Moura},\ and\ \citenamefont
  {Marder}}]{Thompson2009}%
  \BibitemOpen
  \bibfield  {author} {\bibinfo {author} {\bibfnamefont {R.~C.}\ \bibnamefont
  {Thompson-Flagg}}, \bibinfo {author} {\bibfnamefont {M.~J.}\ \bibnamefont
  {Moura}},\ and\ \bibinfo {author} {\bibfnamefont {M.}~\bibnamefont
  {Marder}},\ }\bibfield  {title} {\bibinfo {title} {Rippling of graphene},\
  }\href@noop {} {\bibfield  {journal} {\bibinfo  {journal} {EPL (Europhysics
  Letters)}\ }\textbf {\bibinfo {volume} {85}},\ \bibinfo {pages} {46002}
  (\bibinfo {year} {2009})}\BibitemShut {NoStop}%
\bibitem [{\citenamefont {Deng}\ and\ \citenamefont {Berry}(2016)}]{Deng2016}%
  \BibitemOpen
  \bibfield  {author} {\bibinfo {author} {\bibfnamefont {S.}~\bibnamefont
  {Deng}}\ and\ \bibinfo {author} {\bibfnamefont {V.}~\bibnamefont {Berry}},\
  }\bibfield  {title} {\bibinfo {title} {Wrinkled, rippled and crumpled
  graphene: an overview of formation mechanism, electronic properties, and
  applications},\ }\href@noop {} {\bibfield  {journal} {\bibinfo  {journal}
  {Materials Today}\ }\textbf {\bibinfo {volume} {19}},\ \bibinfo {pages} {197}
  (\bibinfo {year} {2016})}\BibitemShut {NoStop}%
\bibitem [{\citenamefont {Ahmadpoor}\ \emph {et~al.}(2017)\citenamefont
  {Ahmadpoor}, \citenamefont {Wang}, \citenamefont {Huang},\ and\ \citenamefont
  {Sharma}}]{Ahmadpoor2017}%
  \BibitemOpen
  \bibfield  {author} {\bibinfo {author} {\bibfnamefont {F.}~\bibnamefont
  {Ahmadpoor}}, \bibinfo {author} {\bibfnamefont {P.}~\bibnamefont {Wang}},
  \bibinfo {author} {\bibfnamefont {R.}~\bibnamefont {Huang}},\ and\ \bibinfo
  {author} {\bibfnamefont {P.}~\bibnamefont {Sharma}},\ }\bibfield  {title}
  {\bibinfo {title} {Thermal fluctuations and effective bending stiffness of
  elastic thin sheets and graphene: A nonlinear analysis},\ }\href@noop {}
  {\bibfield  {journal} {\bibinfo  {journal} {Journal of the Mechanics and
  Physics of Solids}\ }\textbf {\bibinfo {volume} {107}},\ \bibinfo {pages}
  {294} (\bibinfo {year} {2017})}\BibitemShut {NoStop}%
\bibitem [{\citenamefont {Hassaini}\ \emph {et~al.}(2019)\citenamefont
  {Hassaini}, \citenamefont {Mordant}, \citenamefont {Miquel}, \citenamefont
  {Krstulovic},\ and\ \citenamefont {D\"uring}}]{Hassaini2019}%
  \BibitemOpen
  \bibfield  {author} {\bibinfo {author} {\bibfnamefont {R.}~\bibnamefont
  {Hassaini}}, \bibinfo {author} {\bibfnamefont {N.}~\bibnamefont {Mordant}},
  \bibinfo {author} {\bibfnamefont {B.}~\bibnamefont {Miquel}}, \bibinfo
  {author} {\bibfnamefont {G.}~\bibnamefont {Krstulovic}},\ and\ \bibinfo
  {author} {\bibfnamefont {G.}~\bibnamefont {D\"uring}},\ }\bibfield  {title}
  {\bibinfo {title} {Elastic weak turbulence: From the vibrating plate to the
  drum},\ }\href {https://doi.org/10.1103/PhysRevE.99.033002} {\bibfield
  {journal} {\bibinfo  {journal} {Phys. Rev. E}\ }\textbf {\bibinfo {volume}
  {99}},\ \bibinfo {pages} {033002} (\bibinfo {year} {2019})}\BibitemShut
  {NoStop}%
\bibitem [{\citenamefont {Landau}\ \emph {et~al.}(1986)\citenamefont {Landau},
  \citenamefont {Lifshitz}, \citenamefont {Kosevich},\ and\ \citenamefont
  {Pitaevskii}}]{LandauLifshitzElasticityBook}%
  \BibitemOpen
  \bibfield  {author} {\bibinfo {author} {\bibfnamefont {L.~D.}\ \bibnamefont
  {Landau}}, \bibinfo {author} {\bibfnamefont {E.~M.}\ \bibnamefont
  {Lifshitz}}, \bibinfo {author} {\bibfnamefont {A.~M.}\ \bibnamefont
  {Kosevich}},\ and\ \bibinfo {author} {\bibfnamefont {L.~P.}\ \bibnamefont
  {Pitaevskii}},\ }\href@noop {} {\emph {\bibinfo {title} {Theory of
  elasticity}}},\ \bibinfo {edition} {3rd}\ ed.,\ Vol.~\bibinfo {volume} {7}\
  (\bibinfo  {publisher} {Elsevier},\ \bibinfo {address} {New York},\ \bibinfo
  {year} {1986})\BibitemShut {NoStop}%
\bibitem [{\citenamefont {D\"uring}\ \emph {et~al.}(2006)\citenamefont
  {D\"uring}, \citenamefont {Josserand},\ and\ \citenamefont
  {Rica}}]{During2006}%
  \BibitemOpen
  \bibfield  {author} {\bibinfo {author} {\bibfnamefont {G.}~\bibnamefont
  {D\"uring}}, \bibinfo {author} {\bibfnamefont {C.}~\bibnamefont
  {Josserand}},\ and\ \bibinfo {author} {\bibfnamefont {S.}~\bibnamefont
  {Rica}},\ }\bibfield  {title} {\bibinfo {title} {Weak turbulence for a
  vibrating plate: Can one hear a kolmogorov spectrum?},\ }\href
  {https://doi.org/10.1103/PhysRevLett.97.025503} {\bibfield  {journal}
  {\bibinfo  {journal} {Phys. Rev. Lett.}\ }\textbf {\bibinfo {volume} {97}},\
  \bibinfo {pages} {025503} (\bibinfo {year} {2006})}\BibitemShut {NoStop}%
\bibitem [{\citenamefont {Boudaoud}\ \emph {et~al.}(2008)\citenamefont
  {Boudaoud}, \citenamefont {Cadot}, \citenamefont {Odille},\ and\
  \citenamefont {Touz\'e}}]{Boudaoud2008}%
  \BibitemOpen
  \bibfield  {author} {\bibinfo {author} {\bibfnamefont {A.}~\bibnamefont
  {Boudaoud}}, \bibinfo {author} {\bibfnamefont {O.}~\bibnamefont {Cadot}},
  \bibinfo {author} {\bibfnamefont {B.}~\bibnamefont {Odille}},\ and\ \bibinfo
  {author} {\bibfnamefont {C.}~\bibnamefont {Touz\'e}},\ }\bibfield  {title}
  {\bibinfo {title} {Observation of wave turbulence in vibrating plates},\
  }\href {https://doi.org/10.1103/PhysRevLett.100.234504} {\bibfield  {journal}
  {\bibinfo  {journal} {Phys. Rev. Lett.}\ }\textbf {\bibinfo {volume} {100}},\
  \bibinfo {pages} {234504} (\bibinfo {year} {2008})}\BibitemShut {NoStop}%
\bibitem [{\citenamefont {Mordant}(2008)}]{Mordant2008}%
  \BibitemOpen
  \bibfield  {author} {\bibinfo {author} {\bibfnamefont {N.}~\bibnamefont
  {Mordant}},\ }\bibfield  {title} {\bibinfo {title} {Are there waves in
  elastic wave turbulence?},\ }\href
  {https://doi.org/10.1103/PhysRevLett.100.234505} {\bibfield  {journal}
  {\bibinfo  {journal} {Phys. Rev. Lett.}\ }\textbf {\bibinfo {volume} {100}},\
  \bibinfo {pages} {234505} (\bibinfo {year} {2008})}\BibitemShut {NoStop}%
\bibitem [{\citenamefont {Cadot}\ \emph {et~al.}(2008)\citenamefont {Cadot},
  \citenamefont {Boudaoud},\ and\ \citenamefont {Touz{\'e}}}]{Cadot2008}%
  \BibitemOpen
  \bibfield  {author} {\bibinfo {author} {\bibfnamefont {O.}~\bibnamefont
  {Cadot}}, \bibinfo {author} {\bibfnamefont {A.}~\bibnamefont {Boudaoud}},\
  and\ \bibinfo {author} {\bibfnamefont {C.}~\bibnamefont {Touz{\'e}}},\
  }\bibfield  {title} {\bibinfo {title} {Statistics of power injection in a
  plate set into chaotic vibration},\ }\href@noop {} {\bibfield  {journal}
  {\bibinfo  {journal} {The European Physical Journal B}\ }\textbf {\bibinfo
  {volume} {66}},\ \bibinfo {pages} {399} (\bibinfo {year} {2008})}\BibitemShut
  {NoStop}%
\bibitem [{\citenamefont {Cobelli}\ \emph {et~al.}(2009)\citenamefont
  {Cobelli}, \citenamefont {Petitjeans}, \citenamefont {Maurel}, \citenamefont
  {Pagneux},\ and\ \citenamefont {Mordant}}]{Cobelli2009}%
  \BibitemOpen
  \bibfield  {author} {\bibinfo {author} {\bibfnamefont {P.}~\bibnamefont
  {Cobelli}}, \bibinfo {author} {\bibfnamefont {P.}~\bibnamefont {Petitjeans}},
  \bibinfo {author} {\bibfnamefont {A.}~\bibnamefont {Maurel}}, \bibinfo
  {author} {\bibfnamefont {V.}~\bibnamefont {Pagneux}},\ and\ \bibinfo {author}
  {\bibfnamefont {N.}~\bibnamefont {Mordant}},\ }\bibfield  {title} {\bibinfo
  {title} {Space-time resolved wave turbulence in a vibrating plate},\ }\href
  {https://doi.org/10.1103/PhysRevLett.103.204301} {\bibfield  {journal}
  {\bibinfo  {journal} {Phys. Rev. Lett.}\ }\textbf {\bibinfo {volume} {103}},\
  \bibinfo {pages} {204301} (\bibinfo {year} {2009})}\BibitemShut {NoStop}%
\bibitem [{\citenamefont {Humbert}\ \emph {et~al.}(2013)\citenamefont
  {Humbert}, \citenamefont {Cadot}, \citenamefont {D{\"u}ring}, \citenamefont
  {Josserand}, \citenamefont {Rica},\ and\ \citenamefont
  {Touz{\'e}}}]{Humbert2013}%
  \BibitemOpen
  \bibfield  {author} {\bibinfo {author} {\bibfnamefont {T.}~\bibnamefont
  {Humbert}}, \bibinfo {author} {\bibfnamefont {O.}~\bibnamefont {Cadot}},
  \bibinfo {author} {\bibfnamefont {G.}~\bibnamefont {D{\"u}ring}}, \bibinfo
  {author} {\bibfnamefont {C.}~\bibnamefont {Josserand}}, \bibinfo {author}
  {\bibfnamefont {S.}~\bibnamefont {Rica}},\ and\ \bibinfo {author}
  {\bibfnamefont {C.}~\bibnamefont {Touz{\'e}}},\ }\bibfield  {title} {\bibinfo
  {title} {Wave turbulence in vibrating plates: the effect of damping},\
  }\href@noop {} {\bibfield  {journal} {\bibinfo  {journal} {EPL (Europhysics
  Letters)}\ }\textbf {\bibinfo {volume} {102}},\ \bibinfo {pages} {30002}
  (\bibinfo {year} {2013})}\BibitemShut {NoStop}%
\bibitem [{\citenamefont {Miquel}\ \emph {et~al.}(2013)\citenamefont {Miquel},
  \citenamefont {Alexakis}, \citenamefont {Josserand},\ and\ \citenamefont
  {Mordant}}]{Miquel2013}%
  \BibitemOpen
  \bibfield  {author} {\bibinfo {author} {\bibfnamefont {B.}~\bibnamefont
  {Miquel}}, \bibinfo {author} {\bibfnamefont {A.}~\bibnamefont {Alexakis}},
  \bibinfo {author} {\bibfnamefont {C.}~\bibnamefont {Josserand}},\ and\
  \bibinfo {author} {\bibfnamefont {N.}~\bibnamefont {Mordant}},\ }\bibfield
  {title} {\bibinfo {title} {Transition from wave turbulence to dynamical
  crumpling in vibrated elastic plates},\ }\href
  {https://doi.org/10.1103/PhysRevLett.111.054302} {\bibfield  {journal}
  {\bibinfo  {journal} {Phys. Rev. Lett.}\ }\textbf {\bibinfo {volume} {111}},\
  \bibinfo {pages} {054302} (\bibinfo {year} {2013})}\BibitemShut {NoStop}%
\bibitem [{\citenamefont {D\"uring}\ \emph {et~al.}(2015)\citenamefont
  {D\"uring}, \citenamefont {Josserand},\ and\ \citenamefont
  {Rica}}]{During2015}%
  \BibitemOpen
  \bibfield  {author} {\bibinfo {author} {\bibfnamefont {G.}~\bibnamefont
  {D\"uring}}, \bibinfo {author} {\bibfnamefont {C.}~\bibnamefont
  {Josserand}},\ and\ \bibinfo {author} {\bibfnamefont {S.}~\bibnamefont
  {Rica}},\ }\bibfield  {title} {\bibinfo {title} {Self-similar formation of an
  inverse cascade in vibrating elastic plates},\ }\href
  {https://doi.org/10.1103/PhysRevE.91.052916} {\bibfield  {journal} {\bibinfo
  {journal} {Phys. Rev. E}\ }\textbf {\bibinfo {volume} {91}},\ \bibinfo
  {pages} {052916} (\bibinfo {year} {2015})}\BibitemShut {NoStop}%
\bibitem [{\citenamefont {D{\"u}ring}\ \emph {et~al.}(2017)\citenamefont
  {D{\"u}ring}, \citenamefont {Josserand},\ and\ \citenamefont
  {Rica}}]{During2017}%
  \BibitemOpen
  \bibfield  {author} {\bibinfo {author} {\bibfnamefont {G.}~\bibnamefont
  {D{\"u}ring}}, \bibinfo {author} {\bibfnamefont {C.}~\bibnamefont
  {Josserand}},\ and\ \bibinfo {author} {\bibfnamefont {S.}~\bibnamefont
  {Rica}},\ }\bibfield  {title} {\bibinfo {title} {Wave turbulence theory of
  elastic plates},\ }\href@noop {} {\bibfield  {journal} {\bibinfo  {journal}
  {Physica D: Nonlinear Phenomena}\ }\textbf {\bibinfo {volume} {347}},\
  \bibinfo {pages} {42} (\bibinfo {year} {2017})}\BibitemShut {NoStop}%
\bibitem [{\citenamefont {D{\"u}ring}\ \emph {et~al.}(2019)\citenamefont
  {D{\"u}ring}, \citenamefont {Josserand}, \citenamefont {Krstulovic},\ and\
  \citenamefont {Rica}}]{During2019}%
  \BibitemOpen
  \bibfield  {author} {\bibinfo {author} {\bibfnamefont {G.}~\bibnamefont
  {D{\"u}ring}}, \bibinfo {author} {\bibfnamefont {C.}~\bibnamefont
  {Josserand}}, \bibinfo {author} {\bibfnamefont {G.}~\bibnamefont
  {Krstulovic}},\ and\ \bibinfo {author} {\bibfnamefont {S.}~\bibnamefont
  {Rica}},\ }\bibfield  {title} {\bibinfo {title} {Strong turbulence for
  vibrating plates: Emergence of a kolmogorov spectrum},\ }\href@noop {}
  {\bibfield  {journal} {\bibinfo  {journal} {Physical Review Fluids}\ }\textbf
  {\bibinfo {volume} {4}},\ \bibinfo {pages} {064804} (\bibinfo {year}
  {2019})}\BibitemShut {NoStop}%
\bibitem [{\citenamefont {Nelson}\ and\ \citenamefont
  {Peliti}(1987)}]{Nelson1987}%
  \BibitemOpen
  \bibfield  {author} {\bibinfo {author} {\bibfnamefont {D.}~\bibnamefont
  {Nelson}}\ and\ \bibinfo {author} {\bibfnamefont {L.}~\bibnamefont
  {Peliti}},\ }\bibfield  {title} {\bibinfo {title} {Fluctuations in membranes
  with crystalline and hexatic order},\ }\href@noop {} {\bibfield  {journal}
  {\bibinfo  {journal} {J. Phys. France}\ }\textbf {\bibinfo {volume} {48}},\
  \bibinfo {pages} {1085} (\bibinfo {year} {1987})}\BibitemShut {NoStop}%
\bibitem [{\citenamefont {Kleinert}\ and\ \citenamefont
  {Schulte-Frohlinde}(2001)}]{Kleinert2001}%
  \BibitemOpen
  \bibfield  {author} {\bibinfo {author} {\bibfnamefont {H.}~\bibnamefont
  {Kleinert}}\ and\ \bibinfo {author} {\bibfnamefont {V.}~\bibnamefont
  {Schulte-Frohlinde}},\ }\href@noop {} {\emph {\bibinfo {title} {Critical
  properties of phi4-theories}}}\ (\bibinfo  {publisher} {World Scientific},\
  \bibinfo {address} {Singapore},\ \bibinfo {year} {2001})\BibitemShut
  {NoStop}%
\bibitem [{\citenamefont {McComb}(2003)}]{McComb2003}%
  \BibitemOpen
  \bibfield  {author} {\bibinfo {author} {\bibfnamefont {W.~D.}\ \bibnamefont
  {McComb}},\ }\href@noop {} {\emph {\bibinfo {title} {Renormalization methods:
  a guide for beginners}}}\ (\bibinfo  {publisher} {Oxford University Press},\
  \bibinfo {address} {Oxford, England},\ \bibinfo {year} {2003})\BibitemShut
  {NoStop}%
\bibitem [{\citenamefont {Schwartz}\ and\ \citenamefont
  {Edwards}(1992)}]{Schwartz1992}%
  \BibitemOpen
  \bibfield  {author} {\bibinfo {author} {\bibfnamefont {M.}~\bibnamefont
  {Schwartz}}\ and\ \bibinfo {author} {\bibfnamefont {S.}~\bibnamefont
  {Edwards}},\ }\bibfield  {title} {\bibinfo {title} {Nonlinear deposition: a
  new approach},\ }\href@noop {} {\bibfield  {journal} {\bibinfo  {journal}
  {EPL (Europhysics Letters)}\ }\textbf {\bibinfo {volume} {20}},\ \bibinfo
  {pages} {301} (\bibinfo {year} {1992})}\BibitemShut {NoStop}%
\bibitem [{\citenamefont {Schwartz}\ and\ \citenamefont
  {Edwards}(1998)}]{Schwartz1998}%
  \BibitemOpen
  \bibfield  {author} {\bibinfo {author} {\bibfnamefont {M.}~\bibnamefont
  {Schwartz}}\ and\ \bibinfo {author} {\bibfnamefont {S.~F.}\ \bibnamefont
  {Edwards}},\ }\bibfield  {title} {\bibinfo {title} {Peierls-boltzmann
  equation for ballistic deposition},\ }\href
  {https://doi.org/10.1103/PhysRevE.57.5730} {\bibfield  {journal} {\bibinfo
  {journal} {Phys. Rev. E}\ }\textbf {\bibinfo {volume} {57}},\ \bibinfo
  {pages} {5730} (\bibinfo {year} {1998})}\BibitemShut {NoStop}%
\bibitem [{\citenamefont {Katzav}\ and\ \citenamefont
  {Schwartz}(1999)}]{Katzav1999}%
  \BibitemOpen
  \bibfield  {author} {\bibinfo {author} {\bibfnamefont {E.}~\bibnamefont
  {Katzav}}\ and\ \bibinfo {author} {\bibfnamefont {M.}~\bibnamefont
  {Schwartz}},\ }\bibfield  {title} {\bibinfo {title} {Self-consistent
  expansion for the kardar-parisi-zhang equation with correlated noise},\
  }\href {https://doi.org/10.1103/PhysRevE.60.5677} {\bibfield  {journal}
  {\bibinfo  {journal} {Phys. Rev. E}\ }\textbf {\bibinfo {volume} {60}},\
  \bibinfo {pages} {5677} (\bibinfo {year} {1999})}\BibitemShut {NoStop}%
\bibitem [{\citenamefont {Katzav}\ and\ \citenamefont
  {Schwartz}(2002)}]{Katzav2002}%
  \BibitemOpen
  \bibfield  {author} {\bibinfo {author} {\bibfnamefont {E.}~\bibnamefont
  {Katzav}}\ and\ \bibinfo {author} {\bibfnamefont {M.}~\bibnamefont
  {Schwartz}},\ }\bibfield  {title} {\bibinfo {title} {Existence of the upper
  critical dimension of the kardar--parisi--zhang equation},\ }\href@noop {}
  {\bibfield  {journal} {\bibinfo  {journal} {Physica A: Statistical Mechanics
  and its Applications}\ }\textbf {\bibinfo {volume} {309}},\ \bibinfo {pages}
  {69} (\bibinfo {year} {2002})}\BibitemShut {NoStop}%
\bibitem [{\citenamefont {Schwartz}\ and\ \citenamefont
  {Edwards}(2002)}]{Schwartz2002}%
  \BibitemOpen
  \bibfield  {author} {\bibinfo {author} {\bibfnamefont {M.}~\bibnamefont
  {Schwartz}}\ and\ \bibinfo {author} {\bibfnamefont {S.}~\bibnamefont
  {Edwards}},\ }\bibfield  {title} {\bibinfo {title} {Stretched exponential in
  non-linear stochastic field theories},\ }\href@noop {} {\bibfield  {journal}
  {\bibinfo  {journal} {Physica A: Statistical Mechanics and its Applications}\
  }\textbf {\bibinfo {volume} {312}},\ \bibinfo {pages} {363} (\bibinfo {year}
  {2002})}\BibitemShut {NoStop}%
\bibitem [{\citenamefont {Katzav}(2002)}]{Katzav2002a}%
  \BibitemOpen
  \bibfield  {author} {\bibinfo {author} {\bibfnamefont {E.}~\bibnamefont
  {Katzav}},\ }\bibfield  {title} {\bibinfo {title} {Self-consistent expansion
  for the molecular beam epitaxy equation},\ }\href
  {https://doi.org/10.1103/PhysRevE.65.032103} {\bibfield  {journal} {\bibinfo
  {journal} {Phys. Rev. E}\ }\textbf {\bibinfo {volume} {65}},\ \bibinfo
  {pages} {032103} (\bibinfo {year} {2002})}\BibitemShut {NoStop}%
\bibitem [{\citenamefont {Katzav}(2003{\natexlab{a}})}]{Katzav2003}%
  \BibitemOpen
  \bibfield  {author} {\bibinfo {author} {\bibfnamefont {E.}~\bibnamefont
  {Katzav}},\ }\bibfield  {title} {\bibinfo {title} {Self-consistent expansion
  results for the nonlocal kardar-parisi-zhang equation},\ }\href
  {https://doi.org/10.1103/PhysRevE.68.046113} {\bibfield  {journal} {\bibinfo
  {journal} {Phys. Rev. E}\ }\textbf {\bibinfo {volume} {68}},\ \bibinfo
  {pages} {046113} (\bibinfo {year} {2003}{\natexlab{a}})}\BibitemShut
  {NoStop}%
\bibitem [{\citenamefont {Katzav}(2003{\natexlab{b}})}]{Katzav2003b}%
  \BibitemOpen
  \bibfield  {author} {\bibinfo {author} {\bibfnamefont {E.}~\bibnamefont
  {Katzav}},\ }\bibfield  {title} {\bibinfo {title} {Growing surfaces with
  anomalous diffusion: Results for the fractal kardar-parisi-zhang equation},\
  }\href@noop {} {\bibfield  {journal} {\bibinfo  {journal} {Phys. Rev. E}\
  }\textbf {\bibinfo {volume} {68}},\ \bibinfo {pages} {031607} (\bibinfo
  {year} {2003}{\natexlab{b}})}\BibitemShut {NoStop}%
\bibitem [{\citenamefont {Katzav}\ and\ \citenamefont
  {Schwartz}(2004{\natexlab{a}})}]{Katzav2004}%
  \BibitemOpen
  \bibfield  {author} {\bibinfo {author} {\bibfnamefont {E.}~\bibnamefont
  {Katzav}}\ and\ \bibinfo {author} {\bibfnamefont {M.}~\bibnamefont
  {Schwartz}},\ }\bibfield  {title} {\bibinfo {title} {Kardar-parisi-zhang
  equation with temporally correlated noise: A self-consistent approach},\
  }\href {https://doi.org/10.1103/PhysRevE.70.011601} {\bibfield  {journal}
  {\bibinfo  {journal} {Phys. Rev. E}\ }\textbf {\bibinfo {volume} {70}},\
  \bibinfo {pages} {011601} (\bibinfo {year} {2004}{\natexlab{a}})}\BibitemShut
  {NoStop}%
\bibitem [{\citenamefont {Katzav}\ and\ \citenamefont
  {Schwartz}(2004{\natexlab{b}})}]{Katzav2004a}%
  \BibitemOpen
  \bibfield  {author} {\bibinfo {author} {\bibfnamefont {E.}~\bibnamefont
  {Katzav}}\ and\ \bibinfo {author} {\bibfnamefont {M.}~\bibnamefont
  {Schwartz}},\ }\bibfield  {title} {\bibinfo {title} {Numerical evidence for
  stretched exponential relaxations in the kardar-parisi-zhang equation},\
  }\href {https://doi.org/10.1103/PhysRevE.69.052603} {\bibfield  {journal}
  {\bibinfo  {journal} {Phys. Rev. E}\ }\textbf {\bibinfo {volume} {69}},\
  \bibinfo {pages} {052603} (\bibinfo {year} {2004}{\natexlab{b}})}\BibitemShut
  {NoStop}%
\bibitem [{\citenamefont {Katzav}\ and\ \citenamefont
  {Adda-Bedia}(2006)}]{Katzav2006}%
  \BibitemOpen
  \bibfield  {author} {\bibinfo {author} {\bibfnamefont {E.}~\bibnamefont
  {Katzav}}\ and\ \bibinfo {author} {\bibfnamefont {M.}~\bibnamefont
  {Adda-Bedia}},\ }\bibfield  {title} {\bibinfo {title} {Roughness of tensile
  crack fronts in heterogenous materials},\ }\href@noop {} {\bibfield
  {journal} {\bibinfo  {journal} {EPL (Europhysics Letters)}\ }\textbf
  {\bibinfo {volume} {76}},\ \bibinfo {pages} {450} (\bibinfo {year}
  {2006})}\BibitemShut {NoStop}%
\bibitem [{\citenamefont {Katzav}\ \emph {et~al.}(2007)\citenamefont {Katzav},
  \citenamefont {Adda-Bedia}, \citenamefont {Ben~Amar},\ and\ \citenamefont
  {Boudaoud}}]{Katzav2007}%
  \BibitemOpen
  \bibfield  {author} {\bibinfo {author} {\bibfnamefont {E.}~\bibnamefont
  {Katzav}}, \bibinfo {author} {\bibfnamefont {M.}~\bibnamefont {Adda-Bedia}},
  \bibinfo {author} {\bibfnamefont {M.}~\bibnamefont {Ben~Amar}},\ and\
  \bibinfo {author} {\bibfnamefont {A.}~\bibnamefont {Boudaoud}},\ }\bibfield
  {title} {\bibinfo {title} {Roughness of moving elastic lines: Crack and
  wetting fronts},\ }\href {https://doi.org/10.1103/PhysRevE.76.051601}
  {\bibfield  {journal} {\bibinfo  {journal} {Phys. Rev. E}\ }\textbf {\bibinfo
  {volume} {76}},\ \bibinfo {pages} {051601} (\bibinfo {year}
  {2007})}\BibitemShut {NoStop}%
\bibitem [{\citenamefont {Edwards}\ and\ \citenamefont
  {Schwartz}(2002)}]{Edwards2002}%
  \BibitemOpen
  \bibfield  {author} {\bibinfo {author} {\bibfnamefont {S.~F.}\ \bibnamefont
  {Edwards}}\ and\ \bibinfo {author} {\bibfnamefont {M.}~\bibnamefont
  {Schwartz}},\ }\bibfield  {title} {\bibinfo {title} {Lagrangian statistical
  mechanics applied to non-linear stochastic field equations},\ }\href@noop {}
  {\bibfield  {journal} {\bibinfo  {journal} {Physica A: Statistical Mechanics
  and its Applications}\ }\textbf {\bibinfo {volume} {303}},\ \bibinfo {pages}
  {357} (\bibinfo {year} {2002})}\BibitemShut {NoStop}%
\bibitem [{\citenamefont {Schwartz}\ and\ \citenamefont
  {Katzav}(2008)}]{Schwartz2008}%
  \BibitemOpen
  \bibfield  {author} {\bibinfo {author} {\bibfnamefont {M.}~\bibnamefont
  {Schwartz}}\ and\ \bibinfo {author} {\bibfnamefont {E.}~\bibnamefont
  {Katzav}},\ }\bibfield  {title} {\bibinfo {title} {The ideas behind
  self-consistent expansion},\ }\href@noop {} {\bibfield  {journal} {\bibinfo
  {journal} {Journal of Statistical Mechanics: Theory and Experiment}\ ,\
  \bibinfo {pages} {P04023}} (\bibinfo {year} {2008})}\BibitemShut {NoStop}%
\bibitem [{\citenamefont {Remez}\ and\ \citenamefont
  {Goldstein}(2018)}]{Remez2018}%
  \BibitemOpen
  \bibfield  {author} {\bibinfo {author} {\bibfnamefont {B.}~\bibnamefont
  {Remez}}\ and\ \bibinfo {author} {\bibfnamefont {M.}~\bibnamefont
  {Goldstein}},\ }\bibfield  {title} {\bibinfo {title} {From divergent
  perturbation theory to an exponentially convergent self-consistent
  expansion},\ }\href {https://doi.org/10.1103/PhysRevD.98.056017} {\bibfield
  {journal} {\bibinfo  {journal} {Phys. Rev. D}\ }\textbf {\bibinfo {volume}
  {98}},\ \bibinfo {pages} {056017} (\bibinfo {year} {2018})}\BibitemShut
  {NoStop}%
\bibitem [{\citenamefont {Risken}(1996)}]{RiskenBook}%
  \BibitemOpen
  \bibfield  {author} {\bibinfo {author} {\bibfnamefont {H.}~\bibnamefont
  {Risken}},\ }\href@noop {} {\emph {\bibinfo {title} {The Fokker-Planck
  Equation}}},\ \bibinfo {edition} {2nd}\ ed.\ (\bibinfo  {publisher}
  {Springer},\ \bibinfo {address} {New York},\ \bibinfo {year} {1996})\ pp.\
  \bibinfo {pages} {63--95}\BibitemShut {NoStop}%
\bibitem [{\citenamefont {Balescu}(1975)}]{Balescu1975Book}%
  \BibitemOpen
  \bibfield  {author} {\bibinfo {author} {\bibfnamefont {R.}~\bibnamefont
  {Balescu}},\ }\href {https://books.google.co.il/books?id=5QVRAAAAMAAJ} {\emph
  {\bibinfo {title} {Equilibrium and Non-Equilibrium Statistical Mechanics}}}\
  (\bibinfo  {publisher} {Wiley},\ \bibinfo {address} {New York},\ \bibinfo
  {year} {1975})\BibitemShut {NoStop}%
\bibitem [{\citenamefont {Plischke}\ and\ \citenamefont
  {Bergersen}(2006)}]{PlischkeBergersen1994Book}%
  \BibitemOpen
  \bibfield  {author} {\bibinfo {author} {\bibfnamefont {M.}~\bibnamefont
  {Plischke}}\ and\ \bibinfo {author} {\bibfnamefont {B.}~\bibnamefont
  {Bergersen}},\ }\href@noop {} {\emph {\bibinfo {title} {Equilibrium
  statistical physics}}},\ \bibinfo {edition} {3rd}\ ed.\ (\bibinfo
  {publisher} {World Scientific},\ \bibinfo {address} {Singapore},\ \bibinfo
  {year} {2006})\BibitemShut {NoStop}%
\bibitem [{\citenamefont {Kardar}(2007)}]{Kardar2007Book}%
  \BibitemOpen
  \bibfield  {author} {\bibinfo {author} {\bibfnamefont {M.}~\bibnamefont
  {Kardar}},\ }\href@noop {} {\emph {\bibinfo {title} {Statistical physics of
  particles}}}\ (\bibinfo  {publisher} {Cambridge University Press},\ \bibinfo
  {address} {Cambridge, England},\ \bibinfo {year} {2007})\BibitemShut
  {NoStop}%
\bibitem [{\citenamefont {Poincar\'e}(1893)}]{Poincare1893}%
  \BibitemOpen
  \bibfield  {author} {\bibinfo {author} {\bibfnamefont {H.}~\bibnamefont
  {Poincar\'e}},\ }\href@noop {} {\emph {\bibinfo {title} {Les M\'ethodes
  Nouvelles de la M\'ecanique C\'el\`este, Vol. II}}}\ (\bibinfo  {publisher}
  {Dover},\ \bibinfo {year} {1957 [1893]})\ p.\ \bibinfo {pages}
  {§123–§128}\BibitemShut {NoStop}%
\bibitem [{\citenamefont {Lindstedt}(1882)}]{Lindstedt1882}%
  \BibitemOpen
  \bibfield  {author} {\bibinfo {author} {\bibfnamefont {A.}~\bibnamefont
  {Lindstedt}},\ }\bibfield  {title} {\bibinfo {title} {Beitrag zur integration
  der differentialgleichungen der st\"orungs-theorie},\ }\href@noop {}
  {\bibfield  {journal} {\bibinfo  {journal} {Abh. K. Akad. Wiss. St.
  Petersburg}\ }\textbf {\bibinfo {volume} {31}} (\bibinfo {year}
  {1882})}\BibitemShut {NoStop}%
\bibitem [{\citenamefont {Drazin}(1992)}]{DrazinBook}%
  \BibitemOpen
  \bibfield  {author} {\bibinfo {author} {\bibfnamefont {P.}~\bibnamefont
  {Drazin}},\ }\href@noop {} {\emph {\bibinfo {title} {Nonlinear systems}}}\
  (\bibinfo  {publisher} {Cambridge University Press},\ \bibinfo {address}
  {Cambridge, England},\ \bibinfo {year} {1992})\ p.\ \bibinfo {pages}
  {181–186}\BibitemShut {NoStop}%
\bibitem [{\citenamefont {Katzav}\ and\ \citenamefont
  {Schwartz}(2011{\natexlab{a}})}]{Katzav2011a}%
  \BibitemOpen
  \bibfield  {author} {\bibinfo {author} {\bibfnamefont {E.}~\bibnamefont
  {Katzav}}\ and\ \bibinfo {author} {\bibfnamefont {M.}~\bibnamefont
  {Schwartz}},\ }\bibfield  {title} {\bibinfo {title} {Dynamical inequality in
  growth models},\ }\href {https://doi.org/10.1209/0295-5075/95/66003}
  {\bibfield  {journal} {\bibinfo  {journal} {{EPL} (Europhysics Letters)}\
  }\textbf {\bibinfo {volume} {95}},\ \bibinfo {pages} {66003} (\bibinfo {year}
  {2011}{\natexlab{a}})}\BibitemShut {NoStop}%
\bibitem [{\citenamefont {Katzav}\ and\ \citenamefont
  {Schwartz}(2011{\natexlab{b}})}]{Katzav2011b}%
  \BibitemOpen
  \bibfield  {author} {\bibinfo {author} {\bibfnamefont {E.}~\bibnamefont
  {Katzav}}\ and\ \bibinfo {author} {\bibfnamefont {M.}~\bibnamefont
  {Schwartz}},\ }\bibfield  {title} {\bibinfo {title} {Exponent inequalities in
  dynamical systems},\ }\href {https://doi.org/10.1103/PhysRevLett.107.125701}
  {\bibfield  {journal} {\bibinfo  {journal} {Phys. Rev. Lett.}\ }\textbf
  {\bibinfo {volume} {107}},\ \bibinfo {pages} {125701} (\bibinfo {year}
  {2011}{\natexlab{b}})}\BibitemShut {NoStop}%
\bibitem [{\citenamefont {Hohenberg}\ and\ \citenamefont
  {Halperin}(1977)}]{Hohenberg1977}%
  \BibitemOpen
  \bibfield  {author} {\bibinfo {author} {\bibfnamefont {P.~C.}\ \bibnamefont
  {Hohenberg}}\ and\ \bibinfo {author} {\bibfnamefont {B.~I.}\ \bibnamefont
  {Halperin}},\ }\bibfield  {title} {\bibinfo {title} {Theory of dynamic
  critical phenomena},\ }\href {https://doi.org/10.1103/RevModPhys.49.435}
  {\bibfield  {journal} {\bibinfo  {journal} {Rev. Mod. Phys.}\ }\textbf
  {\bibinfo {volume} {49}},\ \bibinfo {pages} {435} (\bibinfo {year}
  {1977})}\BibitemShut {NoStop}%
\bibitem [{\citenamefont {Ge}\ \emph {et~al.}(2021)\citenamefont {Ge},
  \citenamefont {Meng}, \citenamefont {Song},\ and\ \citenamefont
  {Lam}}]{Ge2021}%
  \BibitemOpen
  \bibfield  {author} {\bibinfo {author} {\bibfnamefont {Z.}~\bibnamefont
  {Ge}}, \bibinfo {author} {\bibfnamefont {N.}~\bibnamefont {Meng}}, \bibinfo
  {author} {\bibfnamefont {L.}~\bibnamefont {Song}},\ and\ \bibinfo {author}
  {\bibfnamefont {E.~Y.}\ \bibnamefont {Lam}},\ }\bibfield  {title} {\bibinfo
  {title} {Dynamic laser speckle analysis using the event sensor},\ }\href
  {https://doi.org/10.1364/AO.412601} {\bibfield  {journal} {\bibinfo
  {journal} {Appl. Opt.}\ }\textbf {\bibinfo {volume} {60}},\ \bibinfo {pages}
  {172} (\bibinfo {year} {2021})}\BibitemShut {NoStop}%
\end{thebibliography}%

\end{document}